\documentclass[a4paper,11pt]{article}
\pdfoutput=1

\usepackage{changes}
\usepackage{jinstpub}
\usepackage{lineno}
\usepackage{booktabs}
\usepackage{subfigure}
\graphicspath{{figure/}}
\usepackage{graphicx}
\usepackage{adjustbox}

\title{Design, construction, and testing of the PandaX-xT cryogenics system}
\author[a]{Xu Wang,}
\author[a,b,*]{Li Zhao,}
\author[d,*]{Xiang Xiao,}
\author[a,b]{Xiangyi,Cui}
\author[a,e]{Shuaijie Li,}
\author[a,b,c]{Jianglai Liu,}

\affiliation[a]{Tsung-Dao Lee institute, New Cornerstone Science Laboratory, Shanghai Jiao Tong University, Shanghai, 200240, China}
\affiliation[b]{Shanghai Jiao Tong University Sichuan Research Institute, Chengdu 610213, China}
\affiliation[c]{INPAC, School of Physics and Astronomy, Shanghai Jiao Tong University,
        \\Key Laboratory for Particle Astrophysics and Cosmology, Ministry of Education (MoE),
         \\Shanghai Key Laboratory for Particle Physics and Cosmology,Shanghai 200240, China}
\affiliation[d]{School of Physics, Sun Yat-Sen University, Guangzhou 510275, China}  
\affiliation[e]{Yalong River Hydropower Development Company, Ltd., Chengdu, 610051,China}

\emailAdd{zhaoli78@sjtu.edu.cn}
\emailAdd{xiaox93@mail.sysu.edu.cn}

\abstract{
The PandaX-xT is a next-generation experiment with broad scientific goals, including the search for dark matter, Neutrinoless Double Beta Decay, and astrophysical neutrinos, using a dual-phase time projection chamber with about 43 tons of liquid xenon. A new cryogenics system of the PandaX-xT is described in this paper. It is developed to handle large mass of liquid xenon efficiently and safely, including two cooling towers for normal operation and one liquid-nitrogen coil for emergency case. Each cooling tower equipped with an AL600 Gifford-McMahon cryocooler features a 1300 W heater, specifically designed to maintain the cold finger's temperature at the desired setpoint. The performance of the cooling tower and the coil has been tested. The cryogenics system with two cooling towers has achieved about 1900~W cooling power at 178~K. The liquid nitrogen coil provides emergency cooling power of more than 1500~W at liquid xenon temperature. For the prototype of a 1-tonne liquid xenon detector, the fluctuation of xenon saturated vapor pressure remains below 1~kPa over one month, while the pressure is around 210~kPa.}

\keywords{Dark matter; Liquid xenon; Gifford-McMahon cryocooler; Liquid-nitrogen.}
\begin{document}
\maketitle
\flushbottom
%\linenumbers

\section{Introduction}
The PandaX experiment, situated at the China Jinping Underground Laboratory (CJPL)~\cite{Kang:2010zza}, employs a dual-phase xenon time projection chamber to search for dark matter particles~\cite{PandaX:2014mjx,PandaX:2015xx,PandaX-II:2016at,PandaX-II:2017xyc,
Zhao:2020zy,PandaX-4T:2023yxn}, specifically weakly interacting massive particles~\cite{PandaX-II:2017cbf,PandaX-4T:2021ym}, and to investigate neutrinoless double beta decay~\cite{PandaX-II:2019kxn,PandaX-4T:2024xyy,PandaX-4T:2025sz}. The use of a larger liquid xenon (LXe) target significantly improves the experimental sensitivity by further suppressing backgrounds. Currently, the PandaX-4T experiment is running with 6 tons of xenon, and the result from the commissioning run and the first science run has been published~\cite{PandaX-4T:2025zhb}. Meanwhile, the XENONnT experiment, utilizing 8 tons of xenon, has also reported new results, and the LZ experiment, with 10 tons of xenon, has delivered the most precise results to date in this field~\cite{XENONnT:2023cxc,XENONnT:2025vwd,LZ:2023lsv,LZ:2024zvo}. The final phase of the PandaX-xT project will feature a detector utilizing 43 tons of xenon~\cite{PandaX-xT:2025xt}. This development will be carried out in two stages: the first, a 20-tonne detector, followed by the 43-tonne one. However, the implementation of this advanced detector will require a new powerful cryogenics system.

In the past few years, several different types of cryogenics systems have been used to support the operation of large LXe detectors. XENONnT has 2 pulse tube refrigerators (PTRs) , the cooling power of each PTR is $\sim$250~W at 177~K~\cite{Aprile:2012jh}. Unlike PTRs of XENONnT, the thermosyphon based cryogenics system with liquid-nitrogen ($LN_{2}$) for LZ experiment, has demonstrated more than 1000~W cooling power at 178~K~\cite{LZ:2017TDR}. The xenon pressure fluctuations during experimental live time are 2~kPa for XENONnT and 0.0558~kPa for LZ~\cite{LZ:2024zvo, XENONnT:2025vwd}. PandaX-4T can run with three cooperating coldheads, the total cooling power is $\sim$580~W at 178~K and the xenon pressure fluctuations, is 0.25~kPa~\cite{Zhao:2020vxh}. Besides, a $LN_{2}$-based cooling system for a next-generation liquid xenon dark matter detector has been proposed~\cite{LN:2020Giboni}. Because of the successful operation of PandaX-4T cryogenics system~\cite{Zhao:2020vxh}, the basic design of PandaX-xT cryogenics system is inherited, with improvement on cooling power, heater of coldhead and control system.

In this paper, we present the results of research and development (R$\&$D) on a new high-performance cryogenics system designed for the future PandaX-xT experiment. 
The system is designed to maintain the inner gas pressure fluctuations below 1~kPa (standard deviation) to ensure stable electroluminescence signals, and achieve the cooling power larger than $\sim$1500~W.
The system integrates two powerful AL600~\cite{GM:AL600} Gifford-McMahon (GM) cryocoolers for long term operation and a specially engineered $LN_{2}$ coil for emergency scenarios. A prototype of the system, featuring a test tower and a 1-tonne liquid xenon detector vessel, demonstrates stable and powerful cooling capabilities. 

\section{Design and construction of the cryogenics system prototype}
\subsection{Overview of the cryogenics system prototype}
The overall heat load breakdown for PandaX-xT experiment is presented in Table~\ref{tab:heat}. Based on the experience from the previous PandaX experiments~\cite{Zhao:2013ghw} and PandaX-4T experiments~\cite{Zhao:2020vxh}, under a vacuum level of $<1\times10^{-3}$~Pa, the heat leak per unit area of the inner vessel is approximately 13~W/m$^2$. This heat leak includes radiation from the outer surface of the inner vessel, feedthrough components, flanges, and signal/HV cables. The first-stage inner vessel (20 tonnes of liquid xenon) of PandaX-xT has a diameter of 2.5 m and a height of 3 m. However, considering future upgrades, we estimate the heat leak based on an inner vessel with a diameter of 3 m and a height of 3.5 m, corresponding to a surface area of 47 m$^2$, resulting in an estimated heat leak of approximately 600 W.
To achieve the required electron lifetime, a circulation flow rate of 500~slpm for purification of 43~tons of liquid xenon is estimated based on experience from previous generations of PandaX experiments. Using a latent heat of vaporization for xenon of 96.3~J/kg and assuming a heat exchanger efficiency of 90$\%$, the resulting heat load is approximately 500 W. Additionally, an all-gas circulation loop with a flow rate of approximately 5~slpm will be employed. 
Xenon gas enters the system at room temperature ($\sim$300~K) and must be cooled to the operating temperature of 178~K, a difference of 122~K. Based on the specific heat capacity of gaseous xenon, the estimated heat load from gas circulation is approximately 9.4~W. Accounting for a safety margin and flow fluctuations, a conservative heat load of 20~W is adopted.
The heat load from the cryogenic pipes, comprising that from the cooling bus and the all-liquid circulation system, totals approximately 180~W. The heat load from the PMTs is dominated by the power dissipation of their bases, which amounts to approximately 25~mW per PMT, while the contribution from the PMT tubes themselves is negligible. Including a contingency of $\sim10\%$, the estimated total heat load is $\sim$1500~W.

\begin{table}[ht]
	\centering
	\caption {Heat load rollup for PandaX-xT experiment, total $\sim$1500~W}
	\label{tab:heat}
	\begin{tabular}[c]{cccc}
		\toprule
		{Component} & {Parameter} & {Heat(W)} & {Note}  \\
		\midrule
		Inner vessel & ID 3~m, H 3.5~m & $\sim$600 & $\sim$13~W/m$^2$ \\
		Heat exchanger & 500~slpm & $\sim$500 & Assumed at 90$\%$ efficiency  \\
	    Gas flow  & 5~slpm  & $\sim$20& Estimated gas circulation  \\
	    Cryogenic pipes & ID 100~mm, ID 35~mm & $\sim$180 & $\sim$20~m, $\sim$80~m\\
	    PMT & 2000 pieces & $\sim$50& $\sim$25~mW/piece  \\
	    Contingency & $\sim10\%$ & $\sim$150 &   \\
		\bottomrule
	\end{tabular}
\end{table}

\begin{figure}[h]
    \centering
    \begin{adjustbox}{center}
        \includegraphics[width=0.75\textheight]{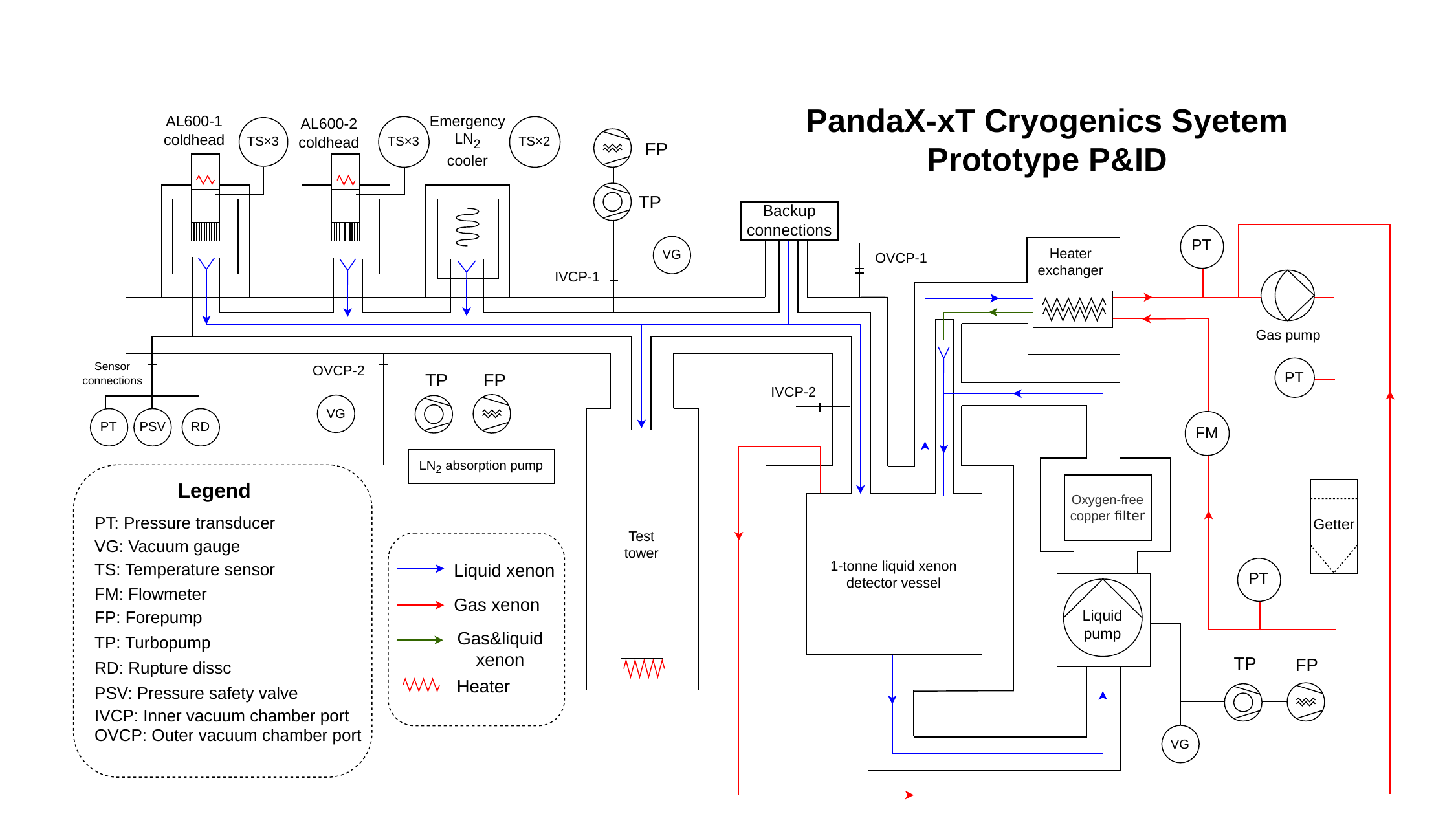}
    \end{adjustbox}
    \caption{Piping and Instrumentation Diagram (P$\&$ID) of the PandaX-xT cryogenics system prototype, illustrating the layout and interconnections of key components including the emergency $LN_{2}$ cooler, two AL600-1 coldheads, 1-tonne liquid xenon detector vessel, a test tower, sensors, and backup connections. This figure also includes a gas circulation system (equipped with a heat exchanger, a gas pump, and a getter), an all-liquid circulation system (equipped with an oxygen-free copper filter and a liquid pump), and an all-gas circulation loop (a stainless steel tube connecting the detector gas space to the gas pump inlet). Additionally, bypass lines and all valves (except for safety valves) are omitted for clarity.}
    \label{fig:schem}
\end{figure}

\begin{figure}[h]
    \centering
    \includegraphics[width = 0.9\textwidth]{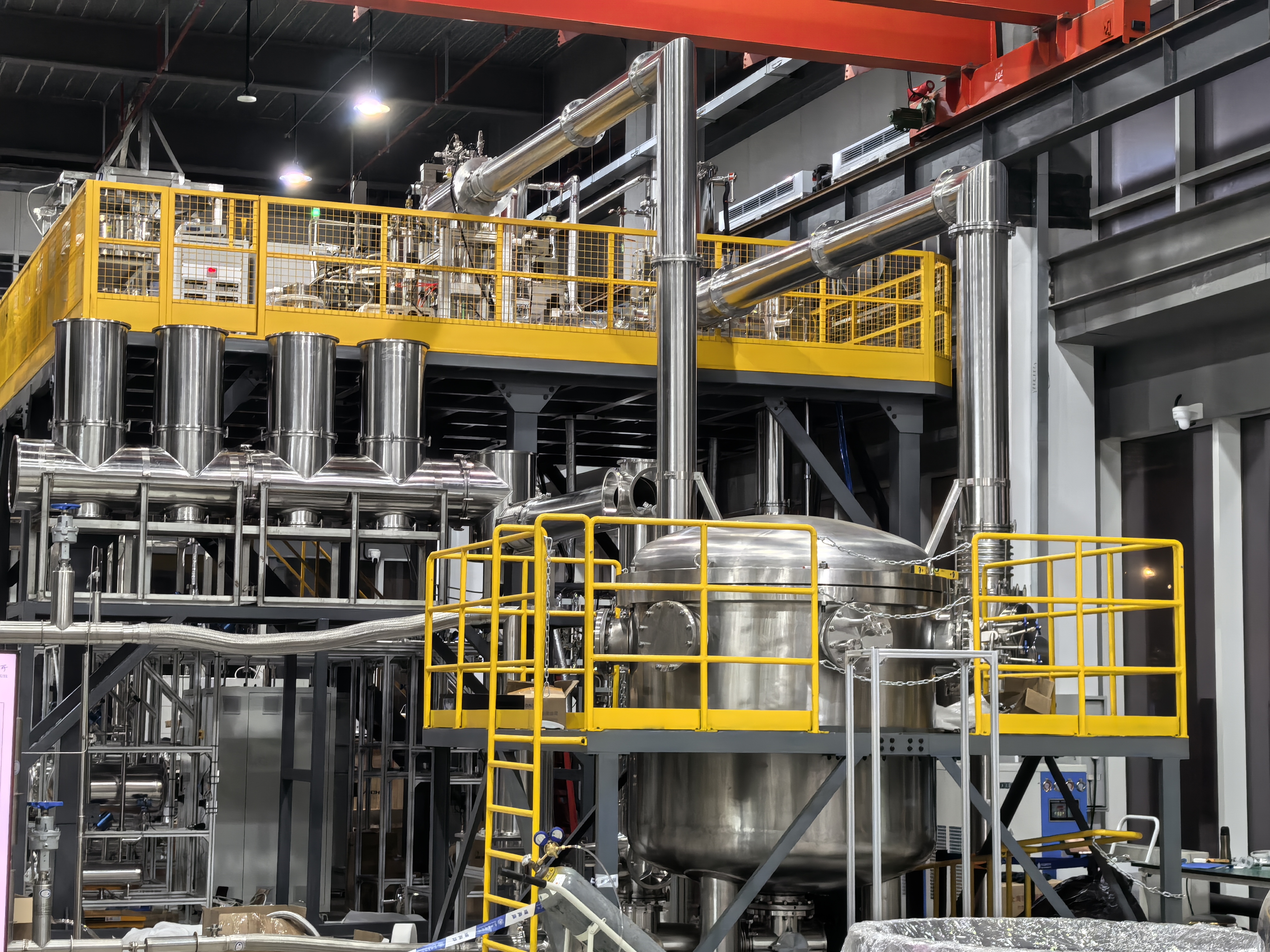}
    \caption{A photograph of the PandaX-xT cryogenics system prototype. The cooling bus, mounted on the upper yellow platform, liquefies xenon gas, which then flows downward by gravity into the 1-tonne liquid xenon detector vessel housed on the lower yellow platform.}
    \label{fig:photo}
\end{figure}

A specially designed cryogenics system prototype, represented schematically in Figure ~\ref{fig:schem} and shown in Figure ~\ref{fig:photo}, mainly includes two cooling towers with AL600 coldhead (Cryomech, USA) with cooling power of 600~W at 80~K and ~1000~W at 178~K~\cite{GM:AL600}, an emergency $LN_{2}$ cooler, a test tower, and a 1-tonne liquid xenon detector vessel. 
Figure~\ref{fig:schem} also illustrates a gas circulation system, an all-liquid circulation system, and an all-gas circulation loop. In practice, four parallel gas circulation systems and four parallel all-gas circulation loops are employed, with only one of each shown for simplicity. The four gas circulation systems, each equipped with a heat exchanger, a gas pump, and a getter, provide a combined flow rate of approximately 500~slpm, ensuring the required purification performance for the PandaX-xT experiment. The all-gas circulation loop extracts gas from the detector and splits into four parallel branches, each feeding into one of the four gas circulation systems for purification. This design is intended to remove impurities from the gas phase space. The all-liquid circulation system, with a designed flow rate of approximately 2~lpm, uses a liquid pump and an oxygen-free copper filter to further remove residual impurities such as O$_2$ in addition to the gas-phase purification, thereby accelerating the overall purification process.
The cooling tower is designed to enable reliable and sustained liquefaction of xenon gas over long operational periods. Detailed parameters of the cooling tower are described in section~\ref{sec:coldhead}. In the event of unexpected power failure or an emergency, $LN_{2}$ is used to maintain the pressure of liquid xenon detector within a safe operational range. The emergency $LN_{2}$ cooler is detailed in section~\ref{sec:LN2}. The test tower, which is equipped with a 2.2~kW heater, is used for simulating high heat load with about 15~kg of liquid xenon. The 1-tonne liquid xenon detector vessel is designed for studying the performance of the liquid xenon detector. Both are described in detail in section~\ref{sec:test-chamber}.

The outer vacuum chamber port-1 is connected to a turbopump (KYKY CXF-250/2301, China) via a 250~mm diameter gate electric valve. The turbopump is backed by a dry screw vacuum pump (Leybold VD65, Germany). This set of pumps is for the insulation and cryostat vacuum. Another turbopump (Leybold 850i, Germany) is mounted for inner chamber port-1 via a metal manual angle valve while inner chamber port-2 is for leak check of the detector vessel. The angle valve would be closed before filling xenon. Its forepump is a dry multi-stage Roots pump (Leybold ECODRY 40 plus, Germany). In addition, the inner pressure sensors, safety valves, rupture discs, and outer gauges are set up at the sensor connections. At the inner pressures exceeding 2.5~barg, the safety valves will open to release xenon to reduce pressure. When the inner pressure exceeds 3~barg, the rupture discs will activate.

Considering the multi-layer insulation (MLI) wrapped onto the inner chamber, the second pumping station with two $LN_{2}$ adsorption pumps is connected to the outer vacuum chamber port-2
directly with a 250~mm diameter empty pipe for maintaining good outer vacuum. The pumping station is also equipped with the pumps, a gate valve, and gauges, which are the same to that of the outer vacuum chamber port-1. The $LN_{2}$ adsorption pump is for emergency cases, such as power-off or malfunctioning of pumps.

Finally, industrial programmable logic system controllers (PLCs) (SIEMENS, SIMATIC S7-1500) are designed and constructed to handle the read-out and the control strategies. For safety, 1~KVA$\ast$24~h uninterruptible power supply (UPS) is set up for PLCs. Figure~\ref{fig:photo} is a photograph of the cryogenics system prototype on the 2nd floor with 1-tonne liquid xenon detector vessel(the test tower behind the vessel is hidden due to the viewing angle) on the 1st floor and two connecting pipes. The facility covers an area of about 100~m$^{2}$.

\subsection{The cooling tower with coldhead}\label{sec:coldhead}
\begin{figure}
    \centering
    \includegraphics[width = 0.8\textwidth]{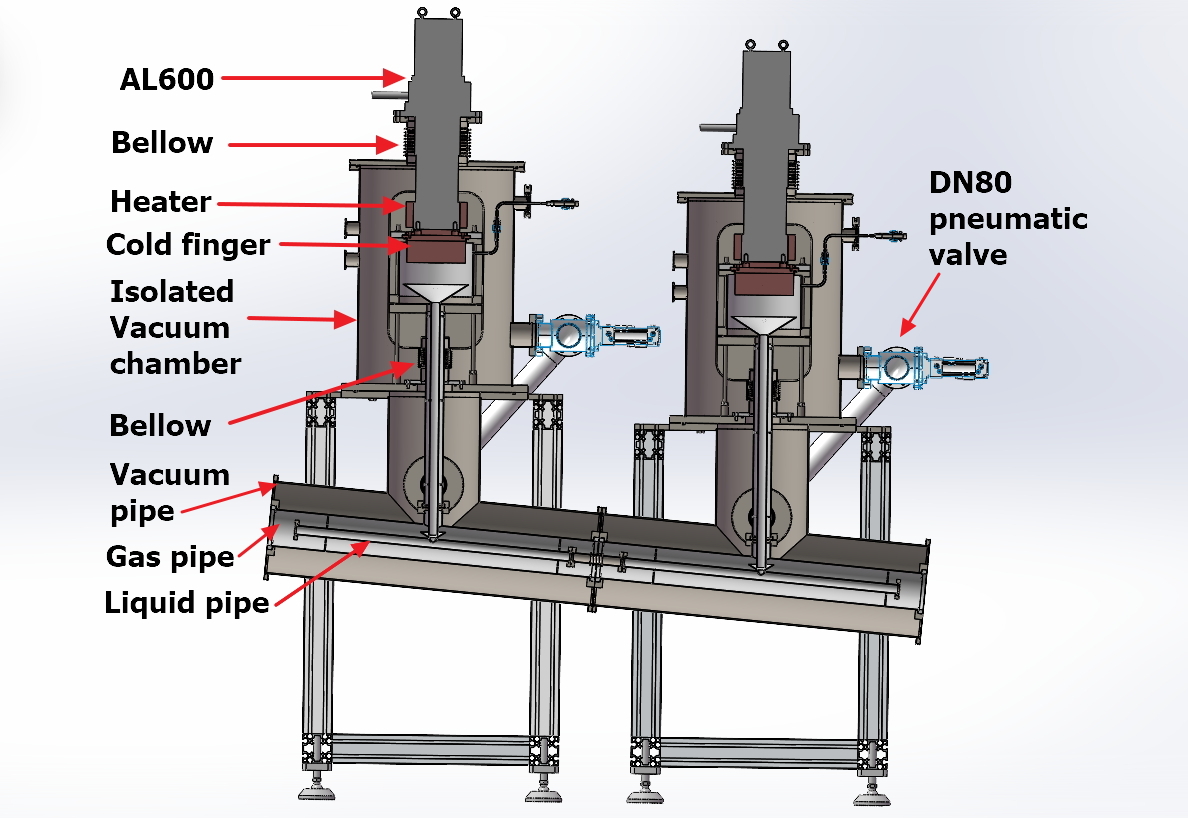}
    \caption{The section view of the cooling tower with AL600 coldhead. The two cooling towers have identical structures. Gas xenon is liquefied on the cold finger and then directed through a funnel into the liquid pipe. The isolated vacuum chamber is controlled by a DN80 pneumatic valve, allowing independent maintenance of each cooling tower without disrupting the operation of the others.}
    \label{fig:coldhead}
\end{figure}

The cooling tower (Figure~\ref{fig:coldhead}) mainly consists of a AL600 coldhead, a heater, a cold finger, and an isolated vacuum chamber. A 165~mm diameter Oxygen-Free High-Conductivity copper disc with the thickness of 16~mm is chosen to be a adapter between the heat exchanger of the coldhead and the cold finger, and they are sealed by flanges with a 0.3 mm thick indium sheet to ensure conductivity for heat between flange connections. The fin of the cold finger is in xenon gas for larger heat exchange area, the drops of liquid xenon condensed on the fin fall into a funnel, which guides them into the central liquid xenon pipe. The liquid xenon flows into the detector through the liquid xenon pipes, transferring cooling power to the detector. The cold finger is sealed with 2~mm diameter indium wire to a 200 mm diameter cylinder vessel, and the inner chamber of this vessel is connected to the gas xenon pipe with a 50 mm diameter pipe.

\begin{figure}
  \centering
  \subfigure[1300~W heater for AL600 coldhead]{\includegraphics[width = 0.4\textwidth]{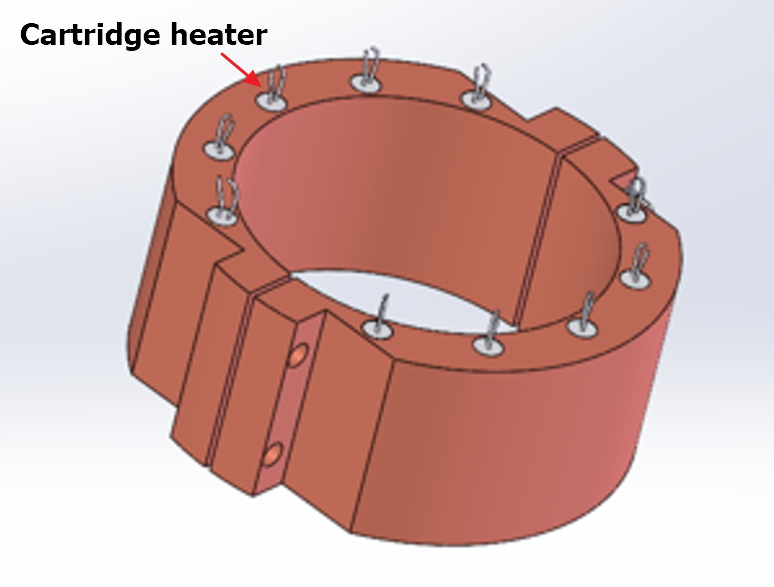}}
  \subfigure[Position of Pt100 sensors]{\includegraphics[width = 0.44\textwidth]{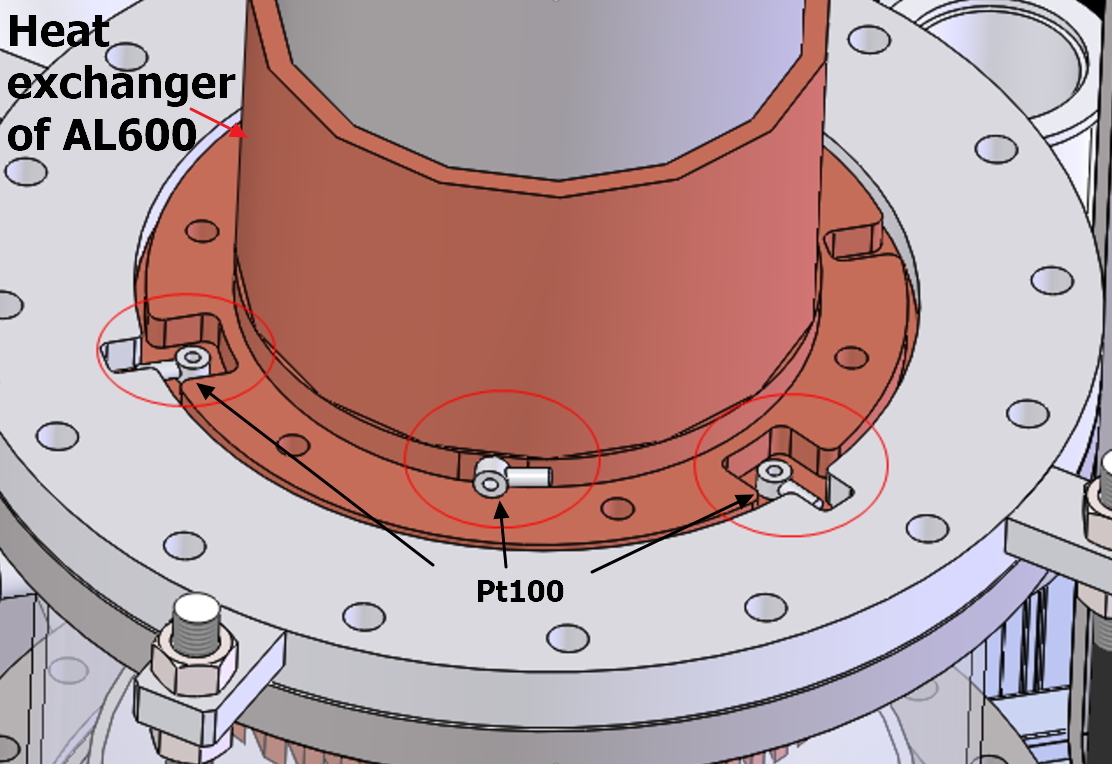}}
  \caption{Design of heater and position of temperature sensors. Cartridge heaters are evenly distributed across the heater to ensure uniform heating, while PT100 sensors are arranged within the same quadrant for ease of installation and replacement.}
  \label{fig:heater-pt100}
\end{figure}

Using cryogenic thermal grease, cartridge heaters (10$\times$130~W) are inserted into suitable holes of the copper ring (Figure~\ref{fig:heater-pt100}.a) composed of two halves, the cooper ring is fixed on the cylinder of the cold head's heat exchanger by screws. Two Pt100 sensors are positioned on the cold finger with screws and another Pt100 sensor is fixed on the copper adapter (Figure~\ref{fig:heater-pt100}.b).
In order to service or replace the coldhead without opening the xenon pipe and breaking the outer vacuum chamber of the detector, the vacuum chamber of the cooling tower is separated from that of the detector with a DB80 pneumatic valve. When the cold head requires replacement, the DN80 valve can be closed to isolate the outer vacuum chamber of cooling tower. This allows the outer vacuum chamber of cooling tower to be vented without breaking the vacuum of the detector. Furthermore, the cold head transfers cooling power to the xenon gas via the copper adapter and cold finger, without direct contact with the xenon. As a result, the integrity of the inner chamber remains preserved during maintenance. The 200~mm diameter inner chamber is supported by four glass fiber reinforced plastics feet, which stand on the bottom plate of the vacuum chamber.

Finally, all the low temperature parts of the cooling tower are enclosed in the 400~mm diameter outer vacuum chamber with 20-layer MLI paper to minimize heat leakage from the outside walls. The vacuum chamber will be connected to the outer vacuum chamber of the detector during the experiment by a DN80 pneumatic valve. The temperature of the cold finger will be regulated by the PLCs and the heater.

\subsection{The emergency $LN_{2}$ cooler}\label{sec:LN2}

\begin{figure}
    \centering
    \includegraphics[width = 0.7\textwidth]{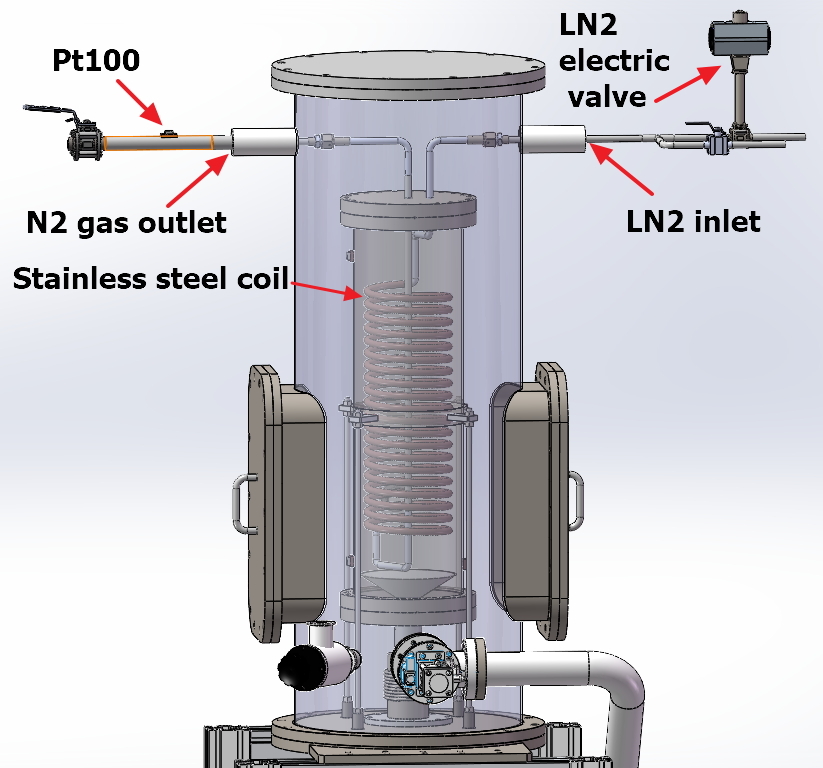}
    \caption{The section view of emergency $LN_{2}$ cooling tower. When the pressure exceeds a preset threshold, the electronic valve opens, allowing $LN_{2}$ to flow into the stainless steel coil. The $LN_{2}$ adsorbs heat from the gas xenon and exits through the outlet. A PT100 temperature sensor monitors the cooling process in real time.}
    \label{fig:LN2}
\end{figure}

The emergency $LN_{2}$ cooler is integrated with a dedicated $LN_{2}$ cooling circuit, an electrically controlled $LN_{2}$ valve, and a Pt100 temperature sensor, as shown in Figure~\ref{fig:LN2}. The cooling medium is provided by one 175~L cryogenic $LN_{2}$ tank with a weight sensor. The $LN_{2}$ cooling circuit is made of a single 9.6~m stainless tube with an inner diameter of 10.22~mm and a wall thickness of 1.24~mm, winded in a circular helix shape. The pitch and the diameter of the circular helix are 23~mm and 152.7~mm, respectively.

The coil dimensions were designed based on previous experience and heat transfer fundamentals~\cite{Incropera:2011}. The dimensions of the coil will be calculated as follows. Taking into account the inner diameter of the coil and the height difference between the liquid nitrogen source and the coil, the flow velocity of liquid nitrogen in the coil is $v = 1.5$~m/s, with a mass flow rate $\dot{m}$ = 99~g/s. The average temperature of the liquid and gaseous nitrogen in the coil is conservatively assumed to be $T_{\text{N}_2}$ = 102~K, while the average temperature of the stainless steel tube wall is taken as $T_{\text{wall}}$ = 105~K. 

Based on those assumptions and mass flow rate, the convective heat transfer coefficient between liquid nitrogen and the stainless steel tube is taken as $h_1$ $\approx$ 257~W/(m$^2\cdot$K), and the natural convection coefficient between xenon gas and the stainless steel tube is $h_2$ $\approx$ 78~W/(m$^2\cdot$K). The thermal conductivity of stainless steel at the average temperature of 105~K is estimated as $k$ $\approx$ 11~W/(m$\cdot$K). 
The overall heat transfer coefficient $h$ is calculated using the thermal resistance network approach~\cite{Incropera:2011}:
\begin{equation}
\frac{1}{h} = \frac{1}{h_1} + \frac{1}{h_2} + \frac{\delta}{k}
\end{equation}
where $\delta$ is the wall thickness of the stainless steel tube. Based on the given parameters, the overall heat transfer coefficient is determined to be:
\begin{equation}
h \approx 59.4 \, \text{W/(m}^2 \cdot \text{K)}.
\end{equation}
To meet the demand for a cooling capacity exceeding 1500\,W, the coil is designed with a cooling power of $Q = 1700$\,W. The required heat transfer area $A$ is then given by
\[
A = \frac{Q}{h \cdot \Delta t} = \frac{1.7 \times 10^3}{59.4 \times (178-102)} \approx 0.38\,\mathrm{m}^{2},
\]
where $\Delta t = 76$\,K is the temperature difference between the inner and outer walls. Given that the outer circumference of the coil is $C = 0.004$\,m, the required coil length $L$ must satisfy
\[
L > \frac{A}{C} = \frac{0.38}{0.004} \approx 9.5\,\mathrm{m}.
\]
Accordingly, a coil with a length of 9.6~m is use in the engineer design, which also provides a small margin.

During failure of the primary AL600 cryocooler due to power loss or system faults, the $LN_{2}$ coil provides intermittent backup cooling via PLC control. In case of PLC control failure during testing, emergency cooling can be activated by manually opening the valve. In the future PandaX-xT experiment, an automated emergency activation system based on the detector pressure and a solenoid valve (on UPS) will be implemented, following the same approach used in the PandaX-4T experiment, thereby eliminating the need for manual intervention. The $LN_{2}$ inlet valve is turned on when the inner pressure (normal, 210~kPa) of the detector is greater than the upper trigger point (230~kPa), and it is turned off when the pressure is less than the lower trigger point (200~kPa). Therefore, the pressure of the liquid xenon detector can be maintained within a safe range. The trigger points also can be set as needed. In addition, a Pt100 temperature sensor at the $N_{2}$ gas outlet is used to monitor the operational status of the emergency $LN_{2}$ cooler. When the emergency cooler is activated, the outlet temperature drops by approximately 100 K, due to the introduction of cold $LN_{2}$ gas.

\subsection{Test tower and 1-tonne liquid xenon vessel}\label{sec:test-chamber}

\begin{figure}
  \centering
  \subfigure[Test tower]{\includegraphics[width = 0.18\textwidth]{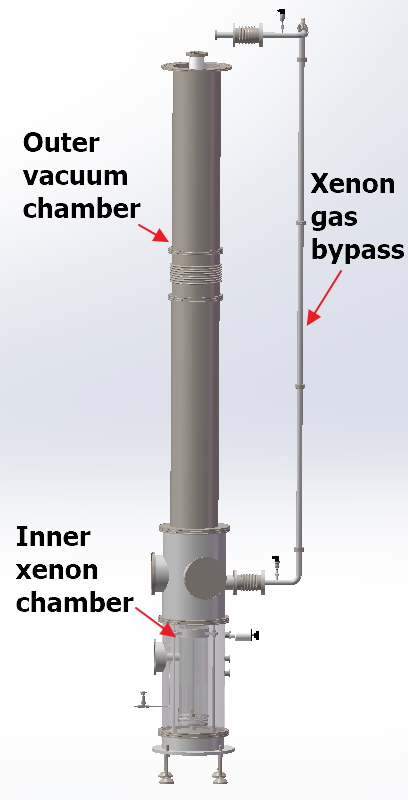}}
  \subfigure[Inner xenon chamber of test tower]{\includegraphics[width = 0.36\textwidth]{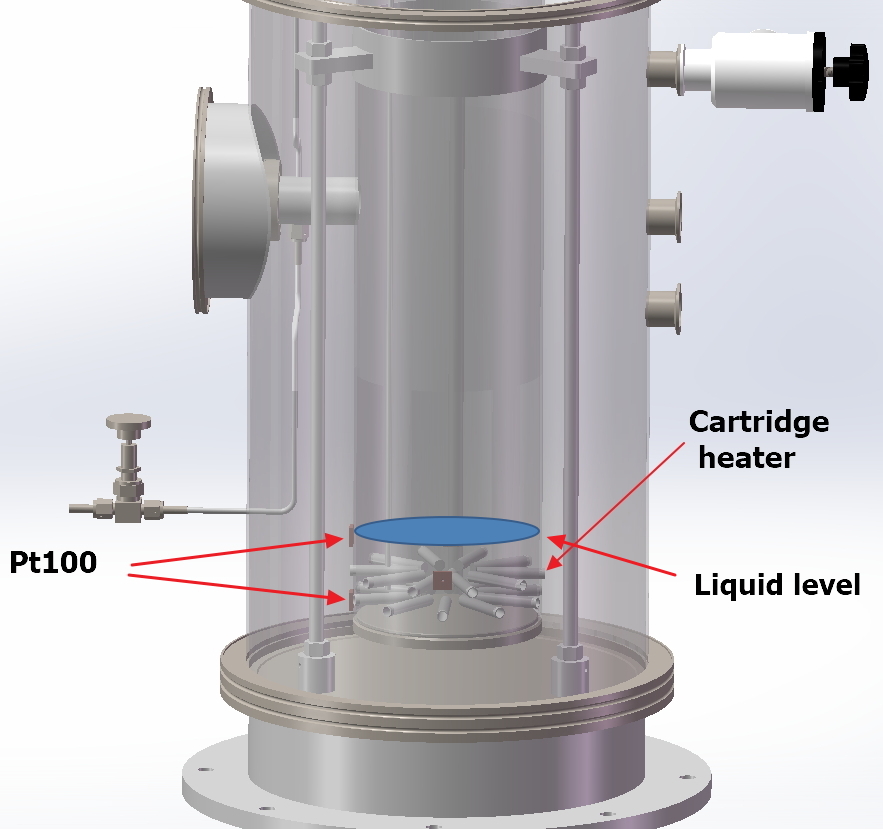}}
  \subfigure[1-tonne liquid xenon detector vessel]{\includegraphics[width = 0.36\textwidth]{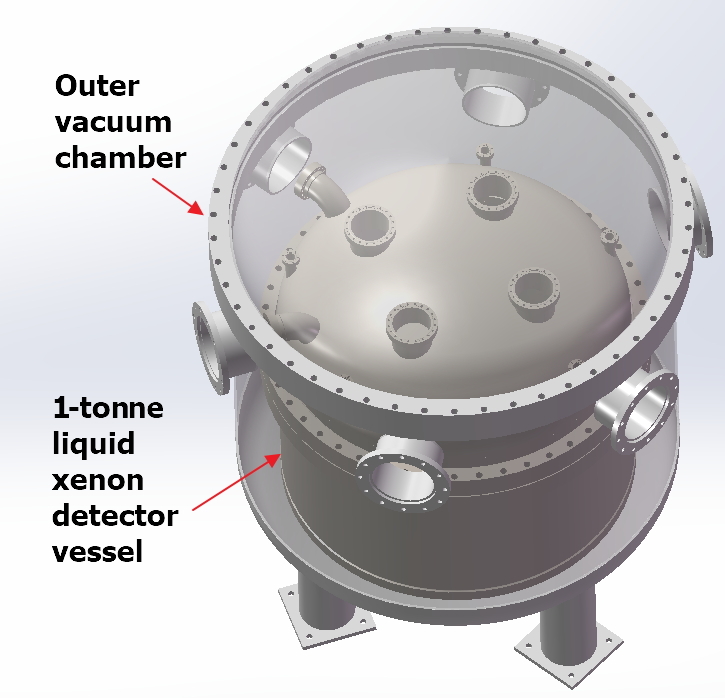}}
  \caption{Test tower and 1-tonne liquid xenon detector vessel. The bottom of the cooling tower is equipped with an inner xenon chamber (15~kg capacity) for testing. At the base of the test tower, an inner xenon chamber (15~kg capacity) is installed for testing. During operation, liquid xenon fills the lower portion of this chamber, with the approximate level marked by a blue solid circle in the figure b. Cartridge heaters assembly, with a total power of 2200~W and mounted at the bottom of the chamber, simulates thermal load to evaluate system stability. The 1-tonne liquid xenon detector vessel is enclosed in an outer vacuum chamber to minimize heat leakage.}
  \label{fig:test-tower-vessel}
\end{figure}

The test tower (Figure~\ref{fig:test-tower-vessel}.a) is designed to test the cooling bus with a small quantity of liquid xenon (about 15~kg) before detector installation. Its main function is to simulate heat leakage from detector, and it features a 2.2~kW heater which exceeds the expected heat load of 1500~W from the future PandaX-xT detector. The bottom of the test tower rests on the ground, its top is connected to the cooling bus on the 2nd floor, and the total height is 4.8~m.
The test tower includes a inner xenon chamber (Figure~\ref{fig:test-tower-vessel}.b) with a diameter of 200~mm and a height of 800~mm. 
To isolate the heaters from direct contact with liquid xenon, blind stainless steel tubes are inserted into pre-drilled holes on the inner xenon chamber wall and hermetically welded to maintain vessel integrity, forming recessed wells for heater installation. Cartridge heaters (20$\times$110~W) are installed within these wells, ensuring efficient heat transfer while preventing xenon contamination. Three Pt100 sensors are mounted at different heights on the outer wall of the inner xenon chamber to monitor both the temperature and liquid level of xenon. All heaters and sensors are housed in the vacuum jacket, with all electrical cables routed within the vacuum. Thus, only the inner surface of the chamber is in direct contact with liquid xenon during operation. The inner xenon chamber is supported by four glass fiber-reinforced plastic feet, which rest on the bottom plate of the outer vacuum chamber. Additionally, a 1-tonne liquid xenon detector vessel (Figure~\ref{fig:test-tower-vessel}.c) was constructed for studying the liquid xenon detector. This vessel has an inner diameter of 1355~mm, a height of 1297~mm, and a flat bottom flange with a diameter of 1373~mm and a thickness of 60~mm. Eight PT100 sensors are installed inside the vessel, with four additional PT100s positioned externally for temperature monitoring.

\section{Experimental results and discussion}
After the system was constructed, at the start of the experiment, the outer and inner chamber were pumped at room temperature. 
Once achieving a high vacuum level ($<1\times10^{-3}$~Pa) in the chamber, a vacuum test was performed on the cooling tower. 
This test aimed to characterize both the effective cooling power and stability of the system. 
The PLCs read the Pt100 sensors in the cold finger and regulate the electrical power supplied to the heaters, keeping the temperature of the cold finger stable at the set value. Subsequently, the effective cooling power will be measured at various temperature setpoints. To assess the system's stability and reliability, each experiment was conducted in cycles (cooling down and warming up) at least three times.

Following the vacumm tests, experiments with liquid xenon were conducted to evaluate the performance of the cooling tower and the emergency $LN_{2}$  cooler.

\begin{figure}
    \centering
    \includegraphics[width = 0.7\textwidth]{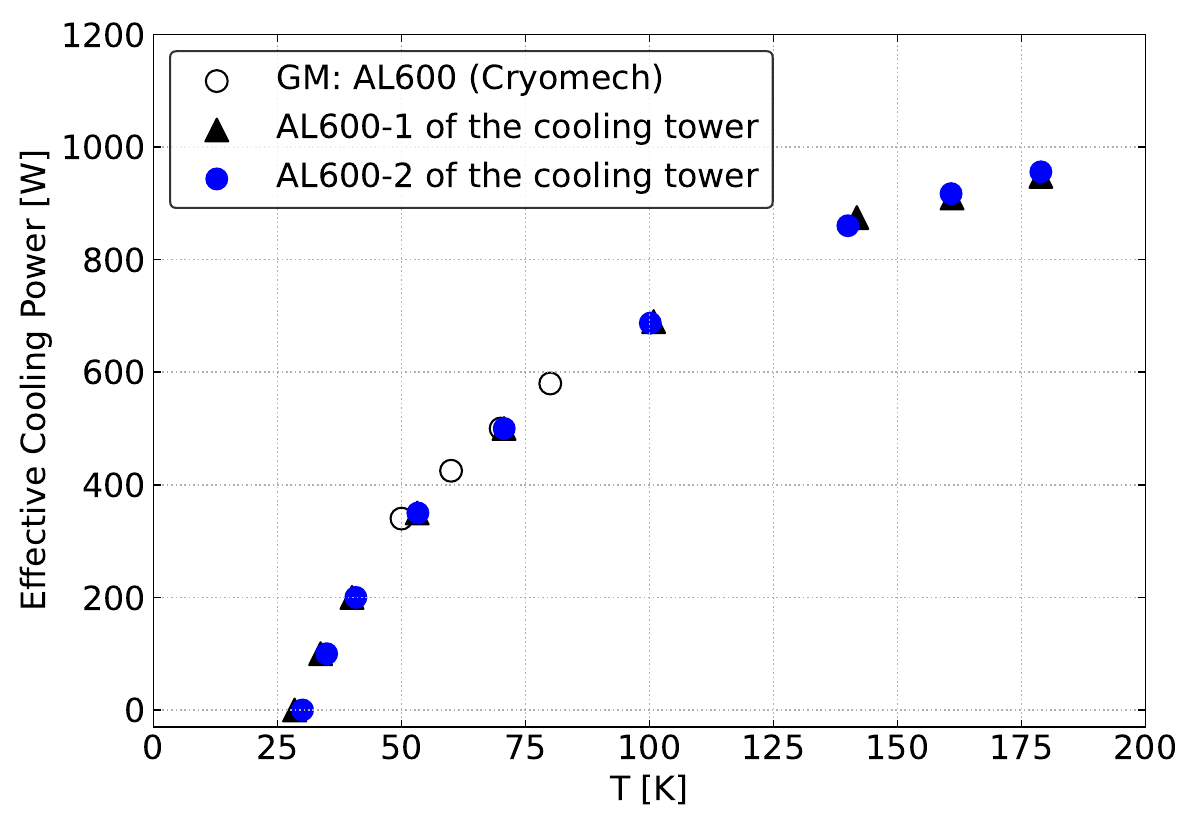}
    \caption{The effective cooling power of the two AL600 cryocoolers is compared with data from Cryomech company. The solid black triangles and solid blue circles represent the actual cooling powers of AL600-1 and AL600-2, respectively. The hollow circles represent four published cooling power points provided by Cryomech.}
    \label{fig:cooling-power-al600}
\end{figure}

\subsection{Vacuum tests of the coldhead AL600}
\begin{figure}
  \centering
  \subfigure[Temperature evolution of AL600-1 ]{\includegraphics[width = 0.45\textwidth]{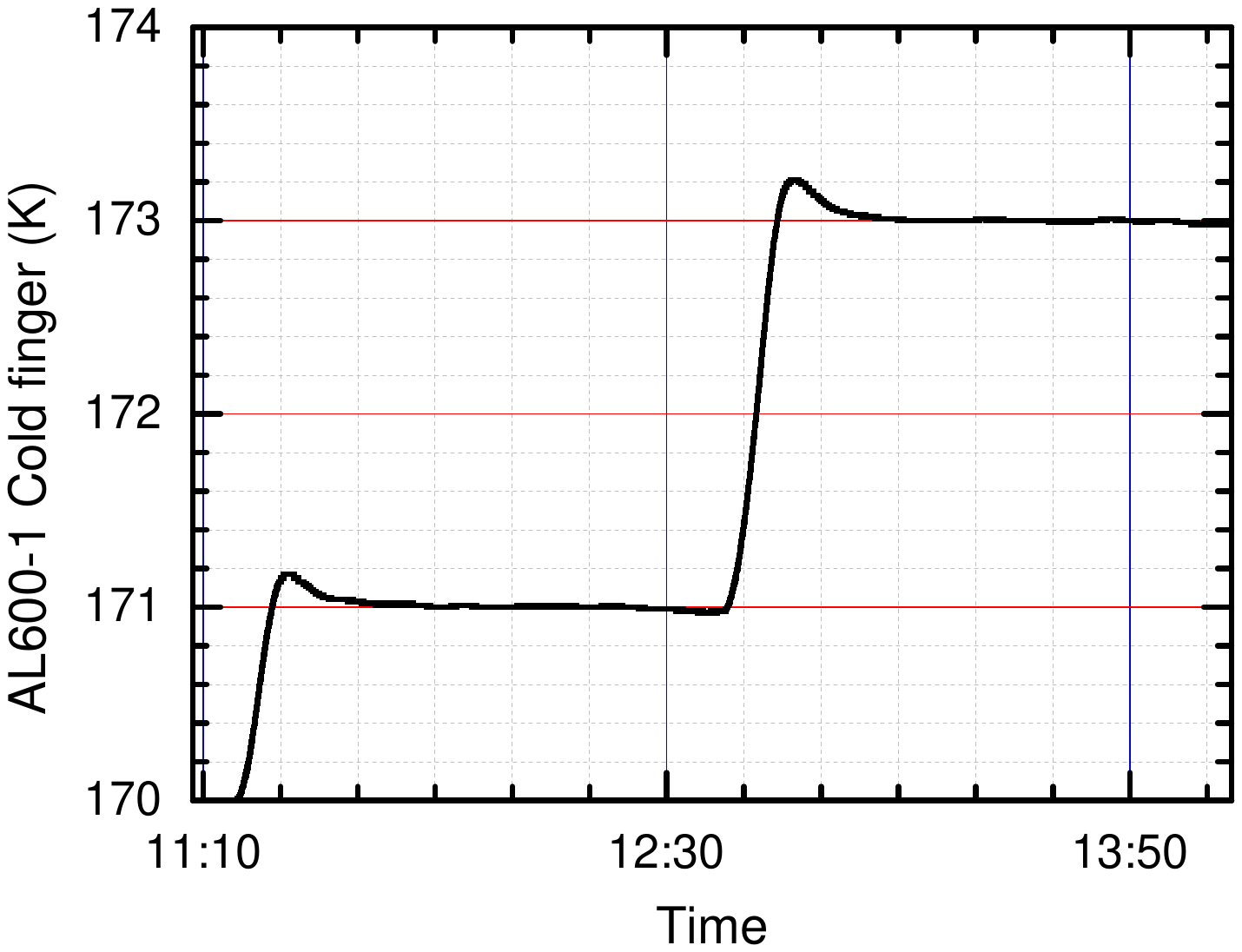}}
  \subfigure[Temperature fluctuation of AL600-1 ]{\includegraphics[width = 0.45\textwidth]{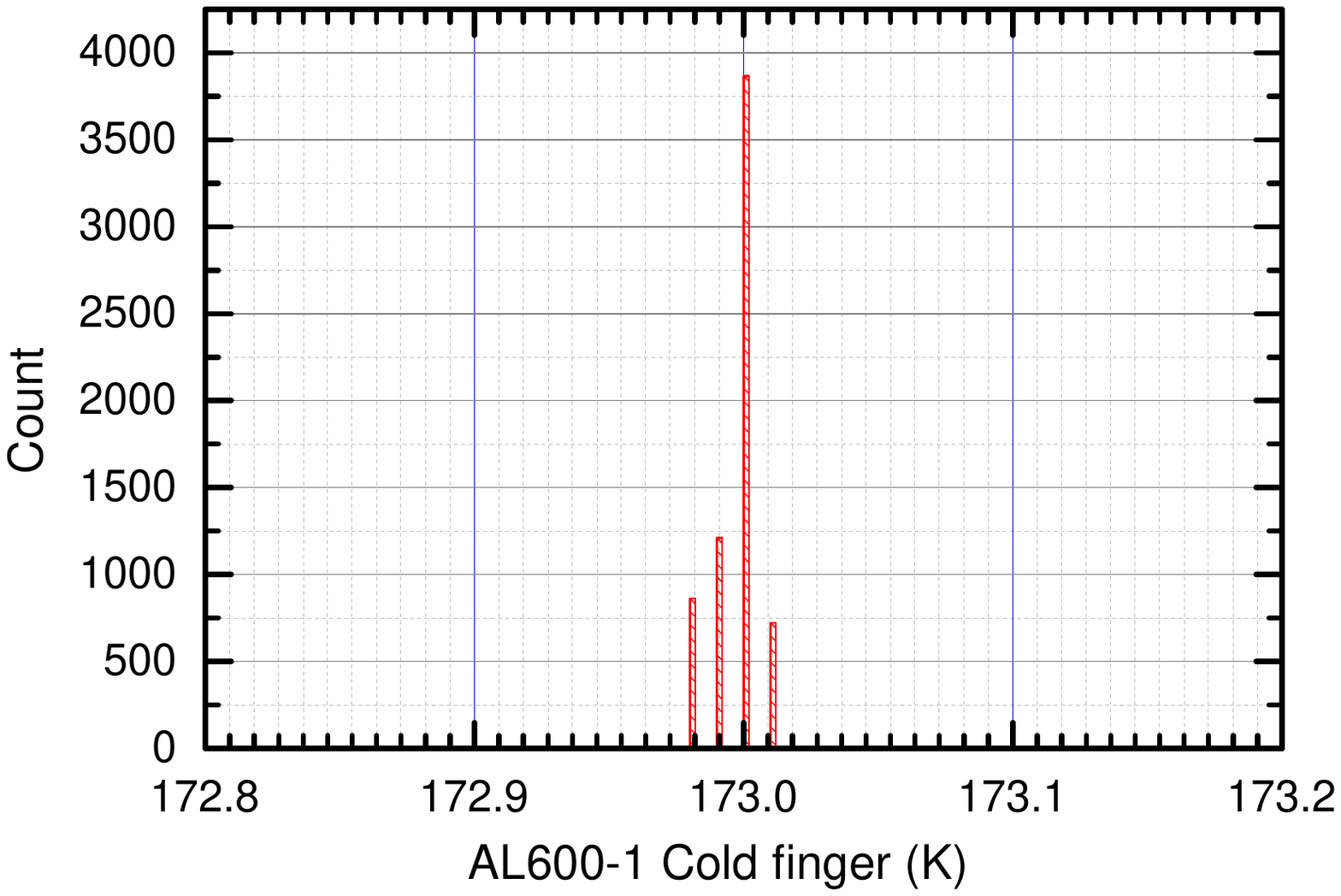}}
  \subfigure[Temperature evolution of AL600-2 ]{\includegraphics[width = 0.45\textwidth]{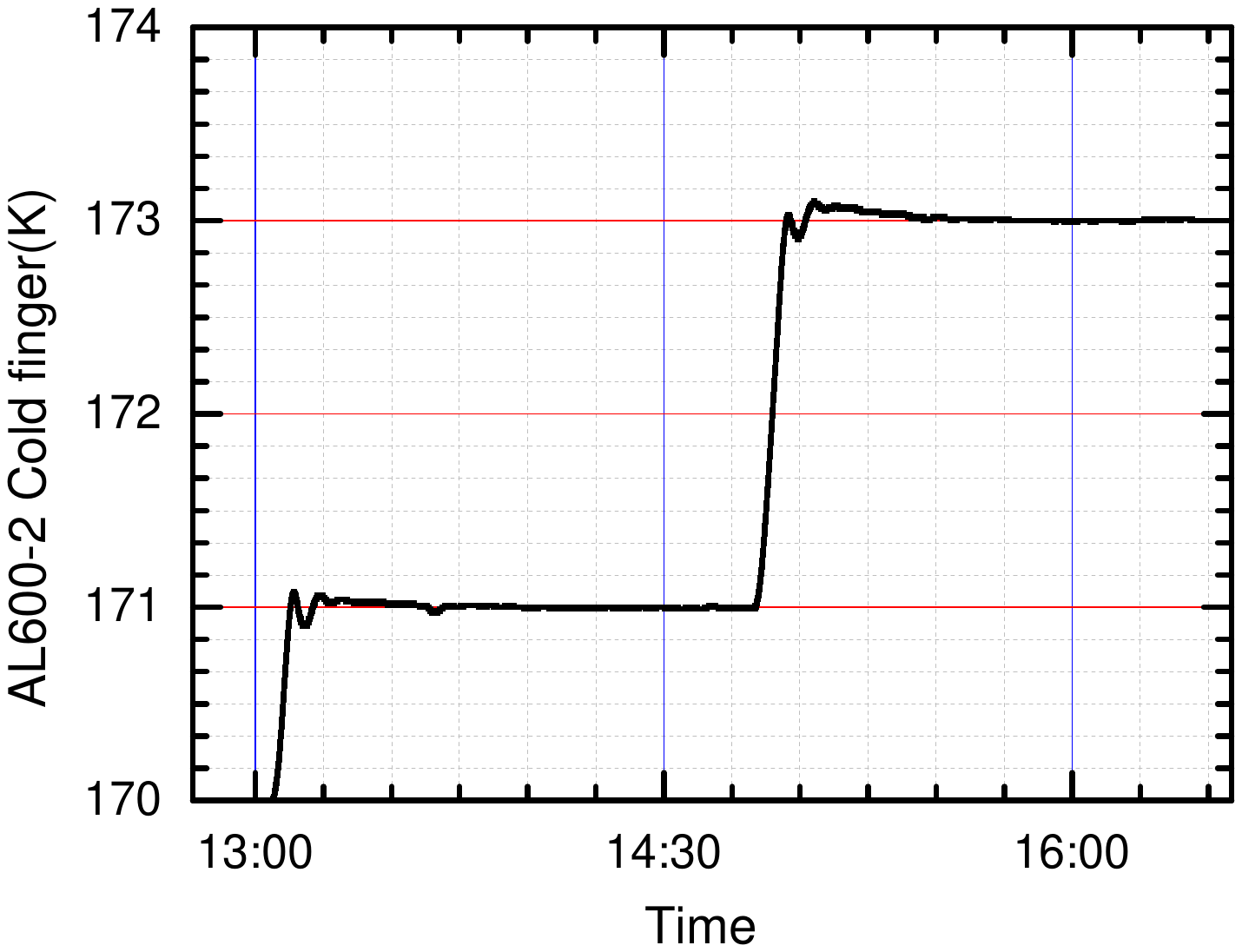}}
  \subfigure[Temperature fluctuation AL600-2 ]{\includegraphics[width = 0.45\textwidth]{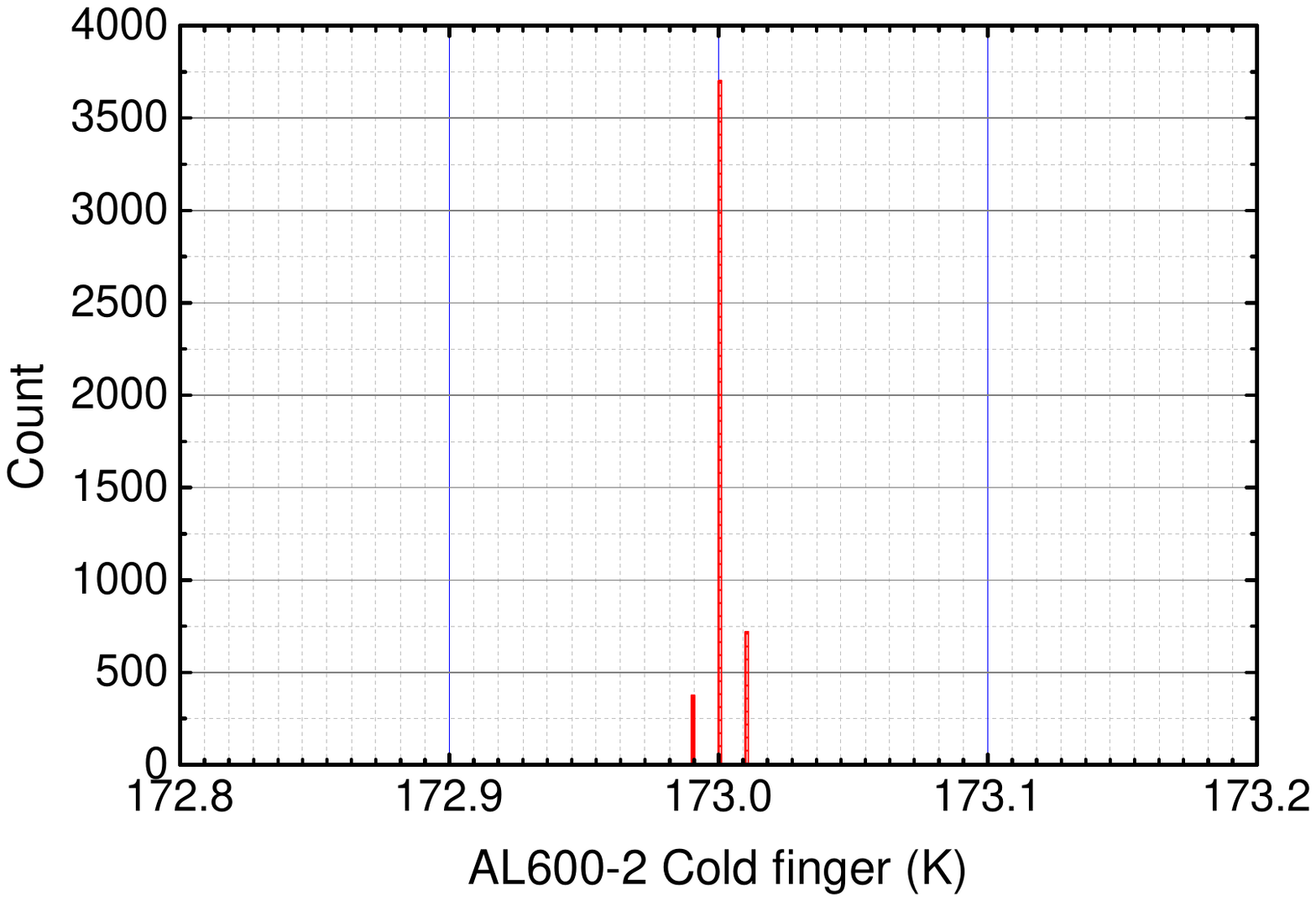}}
  \caption{Temperature of the cold finger and its distribution under vacuum test.Selected temperature evolution from 171~K to 173~K during test, with fluctuation analysis performed on the data collected at the 173~K steady state.}
  \label{fig:temp-al600-1-2}
\end{figure}

The coldheads AL600-1 and AL600-2 were tested under identical vacuum conditions, and their lowest temperatures were measured over multiple runs with zero heater power. After 1.5 hours of cooling, the average minimum temperature reached was $28.60\pm0.13$~K for AL600-1 and $30.12\pm0.13$~K for AL600-2. This corresponds to a temperature difference of $1.52\pm0.18$~K between the two units. Assuming identical intrinsic performance of the two cryocoolers, this difference is attributed to variations in operating environment and installation conditions, particularly the tightness of the thermal interface between the coldhead adapter and the coldfinger, a result that is considered reasonable and acceptable in practical applications.

Nine temperature setpoints, all below 180~K, were tested for each cryocooler to evaluate cooling capacity by heating the cold finger gradually. The temperature of the cold finger stabilizes within 30 minutes after reaching the setpoint, and its fluctuation at steady state is less than 0.05~K. The heating power at steady state is used as the effective cooling power of the coldhead. Figure~\ref{fig:cooling-power-al600} shows the measured cooling capacity of the AL600 cryocooler under different operating temperatures. For reference, the manufacturer-provided cooling power values at 50~K, 60~K, 70~K, and 80~K are also plotted. Although Cryomech only reports data within this limited range, the actual performance of the AL600 in our setup is consistent with the general trend. The cooling power increases with temperature, reaching approximately 950~W at 178~K. The combined cooling capacity of the two coldheads is about 1900~W, which meets the required cooling power.

The temperature evolution at two typical setpoints (171~K and 173~K) and the temperature distribution at 173~K are shown in Figure~\ref{fig:temp-al600-1-2}. When the temperature setpoint was increased from 171~K to 173~K, both coldheads achieved steady-state exceeding 30 minutes, and the temperature fluctuation at 173~K remained below 0.02~K. Despite the similar steady-state performance, the transient responses to setpoint changes differed between the two units (Figure~\ref{fig:temp-al600-1-2}). AL600-1 exhibited a single overshoot, while AL600-2 oscillated before stabilizing. These differences are attributed to the use of different proportional-integral-derivative (PID) tuning parameters: AL600-1 used $\text{P}=3.6$, $\text{I}=960$, $\text{D}=80$, whereas AL600-2 used $\text{P}=5.5$, $\text{I}=960$, $\text{D}=70$. The higher proportional gain in AL600-2 likely contributed to the observed oscillations during the transient phase. The two units were tuned independently during commissioning, which inevitably resulted in different PID parameters for each unit. This is because manufacturing and assembly tolerances throughout the system, including those between the cold head, adapter, and cold finger, as well as between the cold head and the heater, can result in differences in thermal contact and heat transfer efficiency between the two units.

\subsection{Performance of the cooling tower with 800 kg xenon}

\begin{figure}
  \centering
  \subfigure[Xenon pressure over 12 days at 140\,slpm (Run AL600-1)]{\includegraphics[width = 0.48\textwidth]{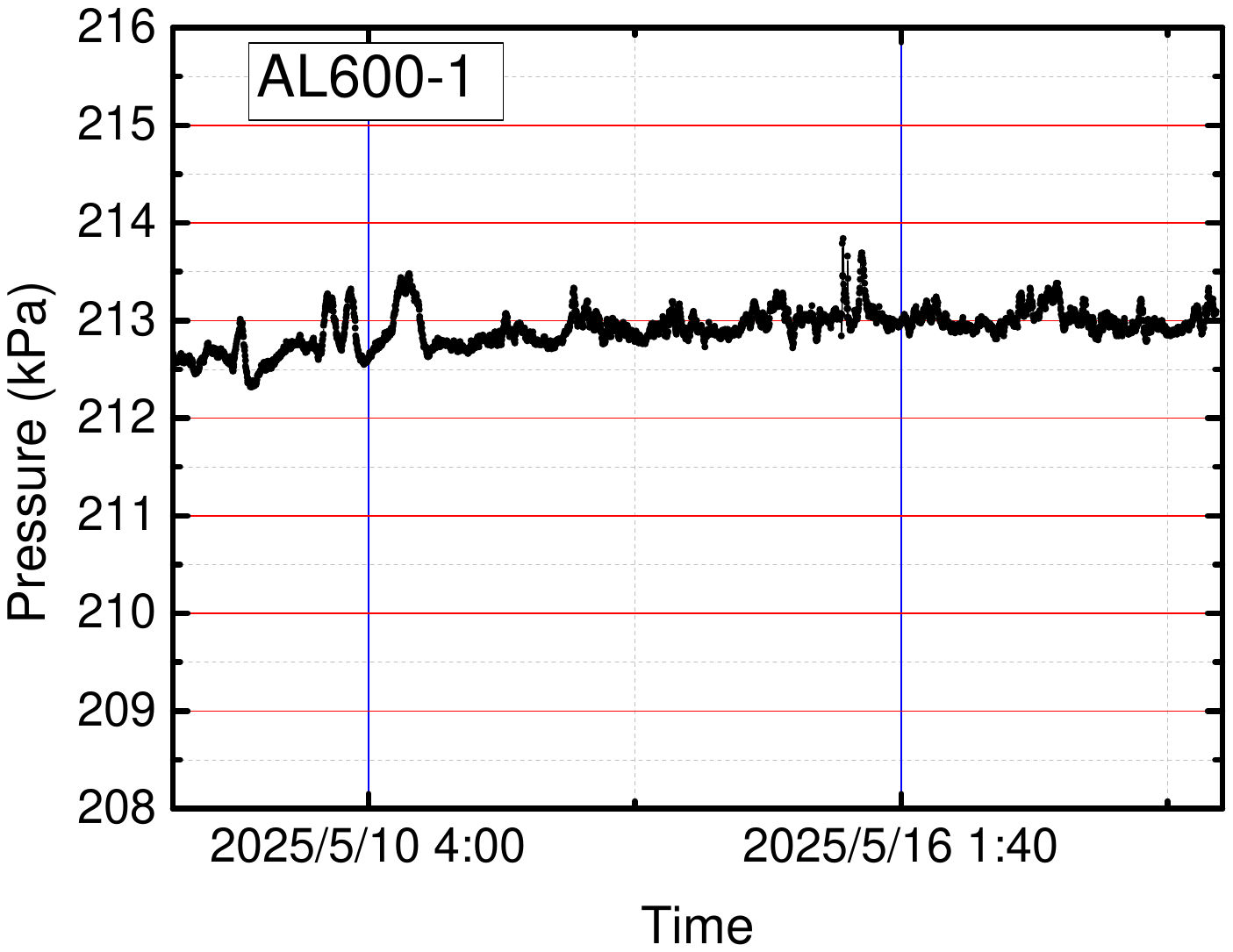}}
  \subfigure[Gaussian fit of pressure fluctuations during Run AL600-1.]{\includegraphics[width = 0.48\textwidth]{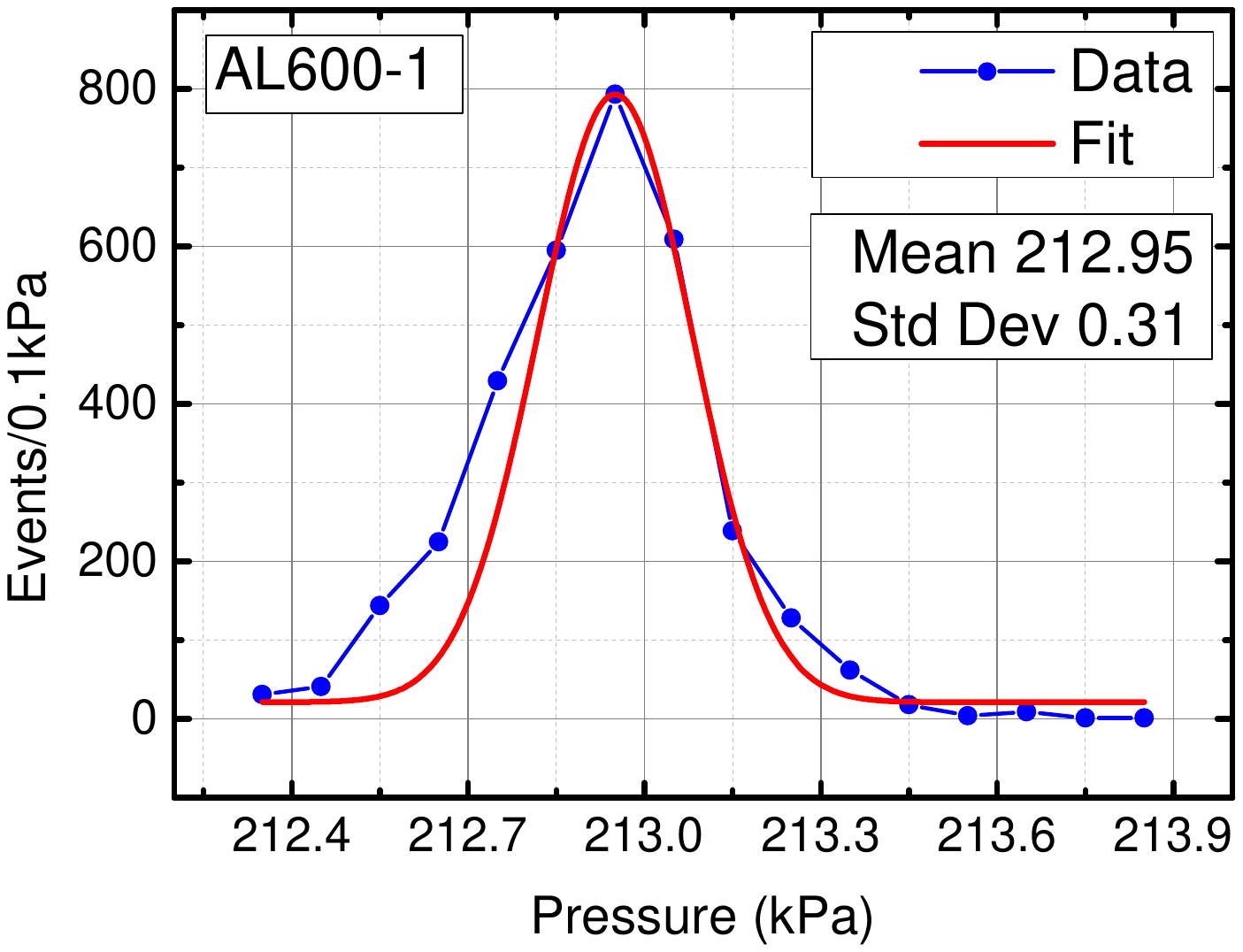}}
  \subfigure[Xenon pressure over 30 days at 80\,slpm (Run AL600-2)]{\includegraphics[width = 0.48\textwidth]{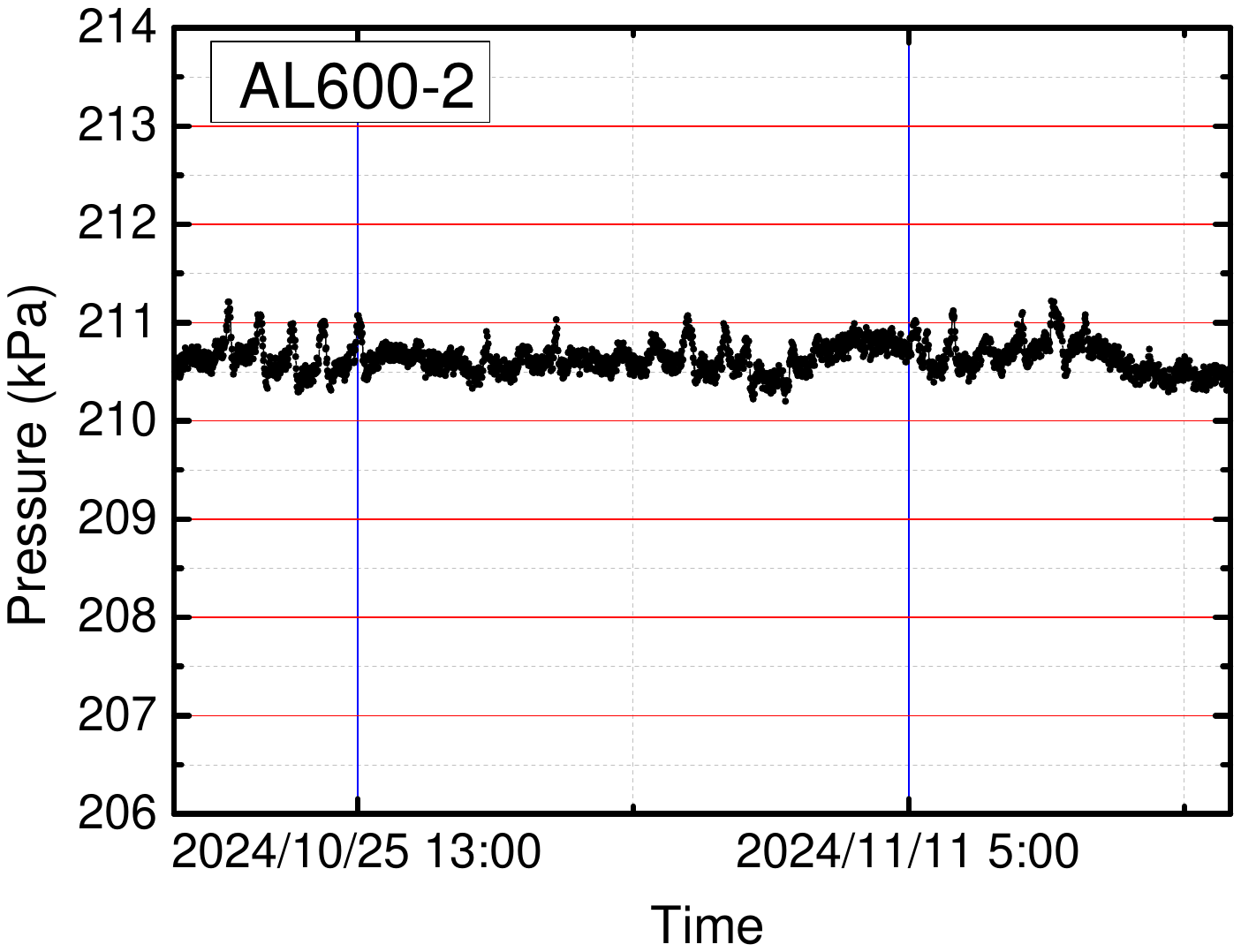}}
  \subfigure[Gaussian fit of pressure fluctuations during Run AL600-2]{\includegraphics[width = 0.48\textwidth]{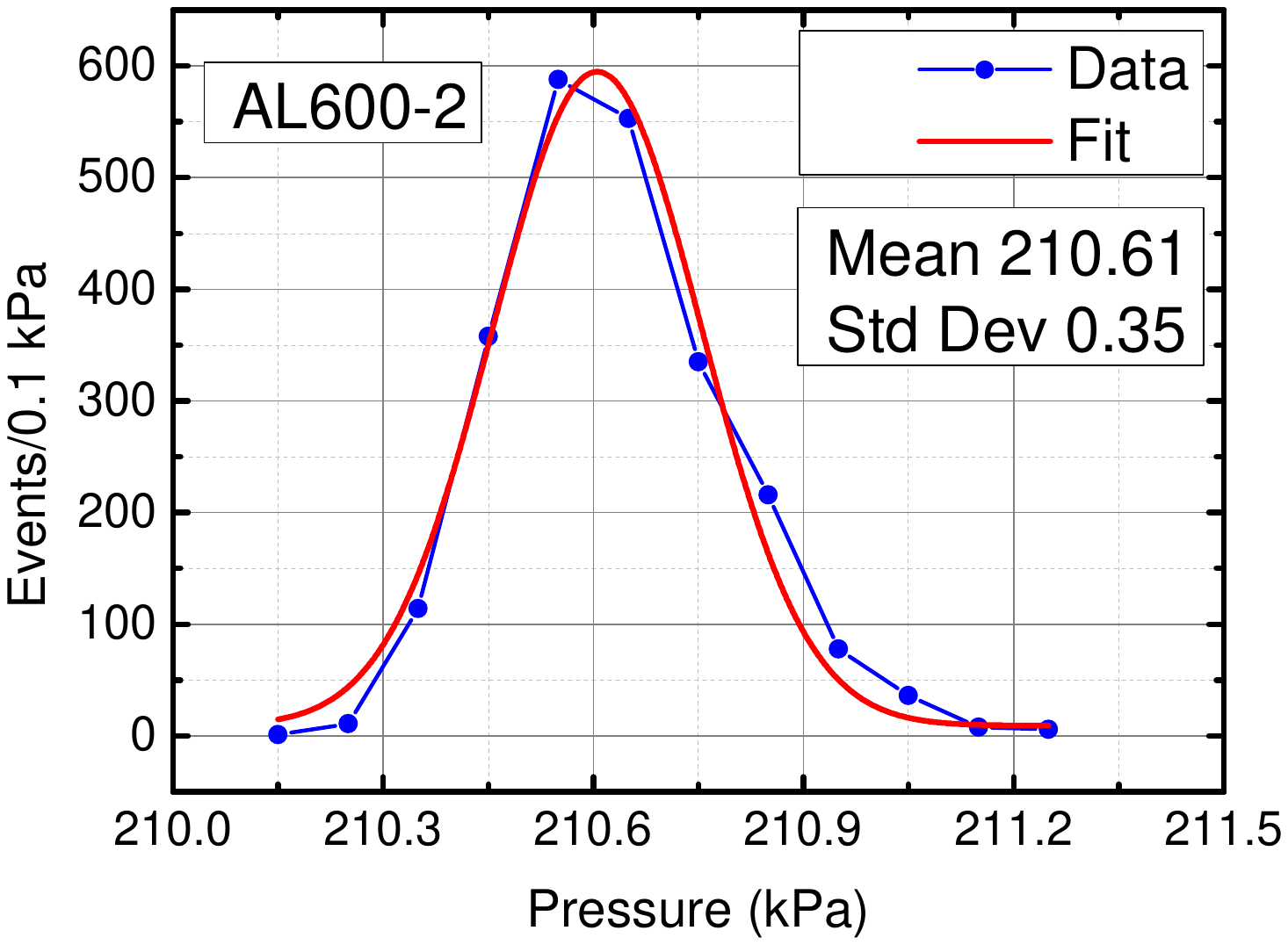}}
  \caption{Performance of the cooling towers under different circulation rates. Shown are the long-term pressure behavior and fluctuation statistics from Runs AL600-1 (140~slpm) and AL600-2 (80~slpm) in the 1-tonne liquid xenon detector.}
  \label{fig:p-al600-1-2}
\end{figure}

The cooling tower was then used to stabilize the temperature and pressure of the 1-tonne liquid xenon detector prototype which contained approximately 800 kg of liquid xenon. At the same time, the online gas circulation systems were in operation. As shown in Figure~\ref{fig:p-al600-1-2}, the xenon pressure remained stable for half a month using the AL600-1 coldhead (cold finger setpoint: 178~K) at a gas purification flow rate of approximately 140~slpm. Similarly, with the AL600-2 coldhead (cold finger setpoint: 178.5~K) and a flow rate of about 80~slpm, due to the replacement of circulation pump, stable pressure was maintained for one month. The pressure data of two cold heads were fitted with a Gaussian function and their standard deviation of the pressure is 0.31~kPa and 0.35~kPa respectively. During independent operation, each cryocooler required a heater power exceeding 650 W to stabilize the system pressure at $212.95\pm0.31$~kPa and $210.61\pm0.35$~kPa, respectively, demonstrating that both units retained an available cooling capacity greater than 650 W at the operating temperature. Furthermore, these two cryocoolers can operate simultaneously to support higher thermal loads. 

The total system heat loads were measured under different gas circulation flow rates. At a flow rate of 80~slpm, the measured heat load was approximately 218~W, which increased to approximately 282~W at 140~slpm. The theoretical model estimates a total heat load of 235~W (at 80~slpm) and 295~W (at 140~slpm), respectively. The heat load composition can be broken down as follows. With a total surface area of 8.5~m$^2$ and a unit area heat leak of 13~W/m$^2$, the inner vessel of the 1-ton liquid xenon detector contributes a static heat load of approximately 110~W. 
The circulation system introduces a flow‑dependent heat load of about 1~W/slpm, assuming a heat exchanger efficiency of 90$\%$.
Under flow rates of 80~slpm and 140~slpm, the resulting heat loads are 80~W and 140~W, respectively. The prototype employs a total of 25~m of cryogenic piping, contributing an estimated static heat leak of approximately 45~W. The all-gas circulation loop and all-liquid circulation system was not activated during the tests, thus contributing zero heat load. The heat contribution from the PMTs was considered negligible as they were only powered on for brief periods. 
The good agreement between the measured and calculated total heat loads confirms our understanding of thermal characteristics of the system.

\begin{figure}
  \centering
  \subfigure[Temperature distribution of AL600-1.]{\includegraphics[width = 0.45\textwidth]{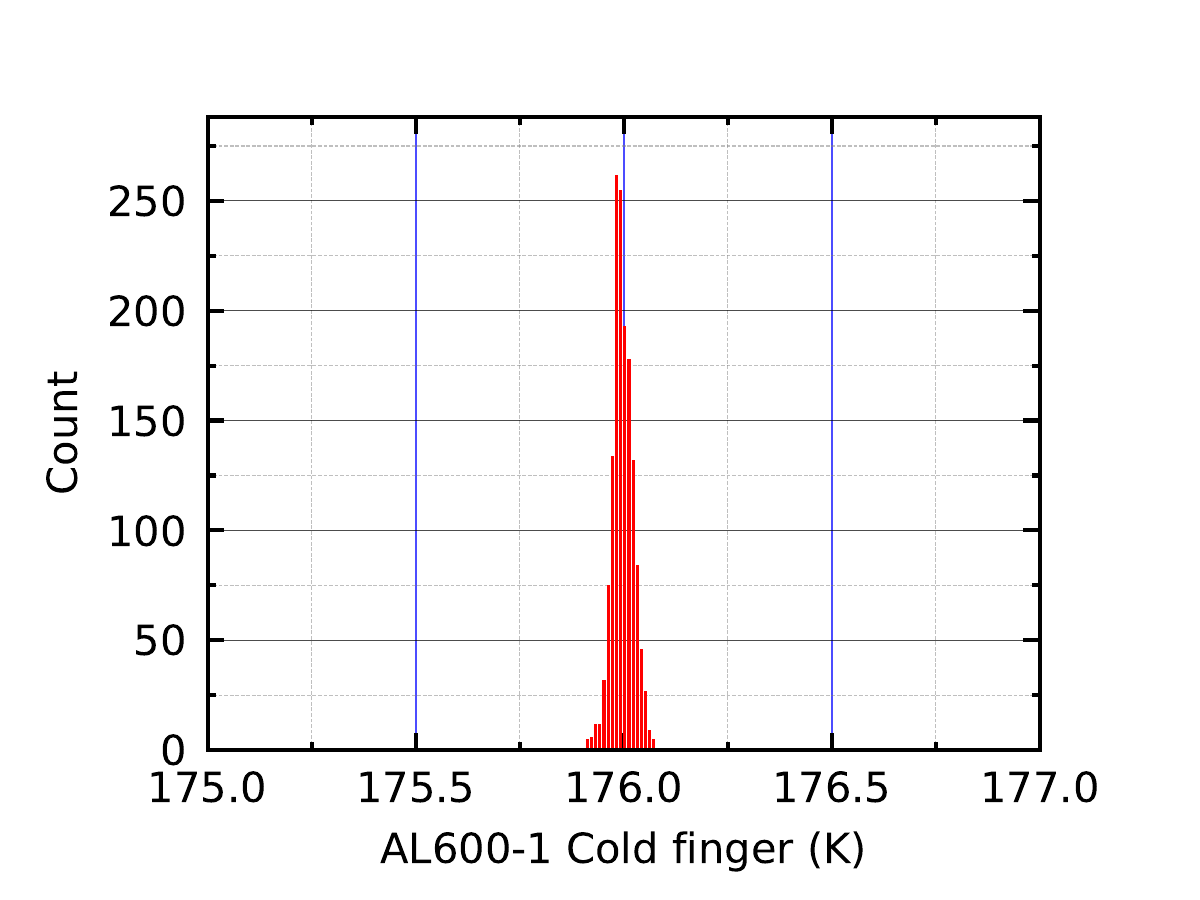}}
  \subfigure[Temperature distribution of AL600-2.]{\includegraphics[width = 0.45\textwidth]{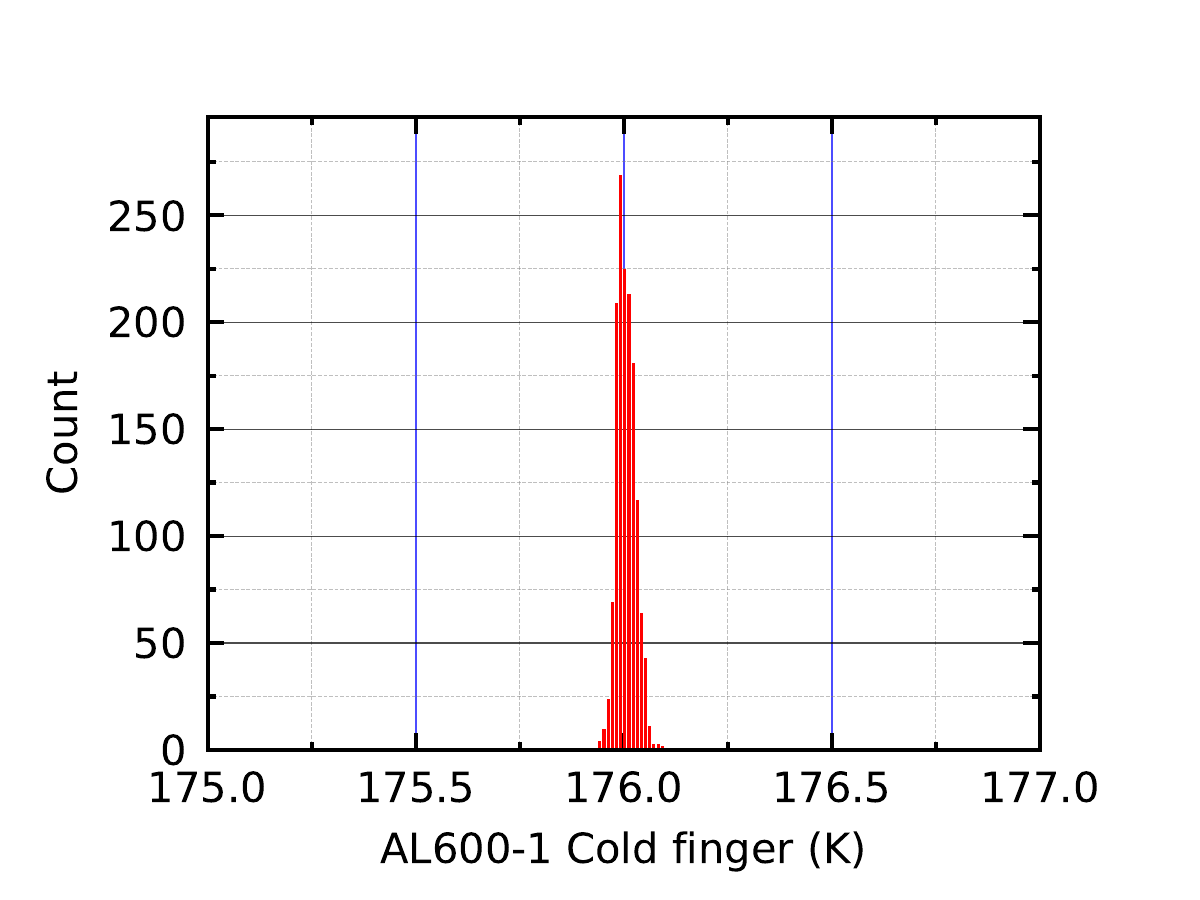}}
  \subfigure[Xenon pressure of one day during two cooling heads operate simultaneously.]{\includegraphics[width = 0.45\textwidth]{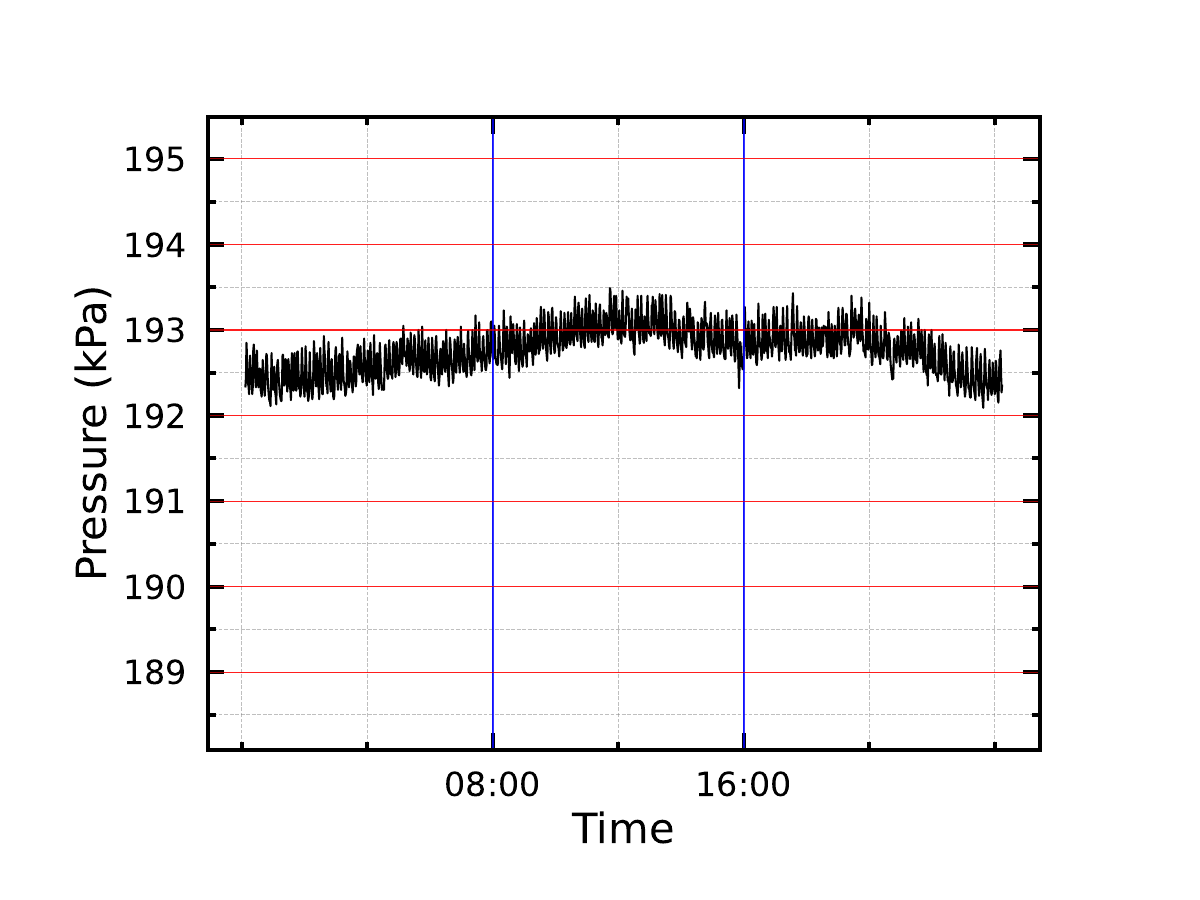}}
  \subfigure[Gaussian fit of pressure fluctuations during two cooling heads operate simultaneously.]{\includegraphics[width = 0.45\textwidth]{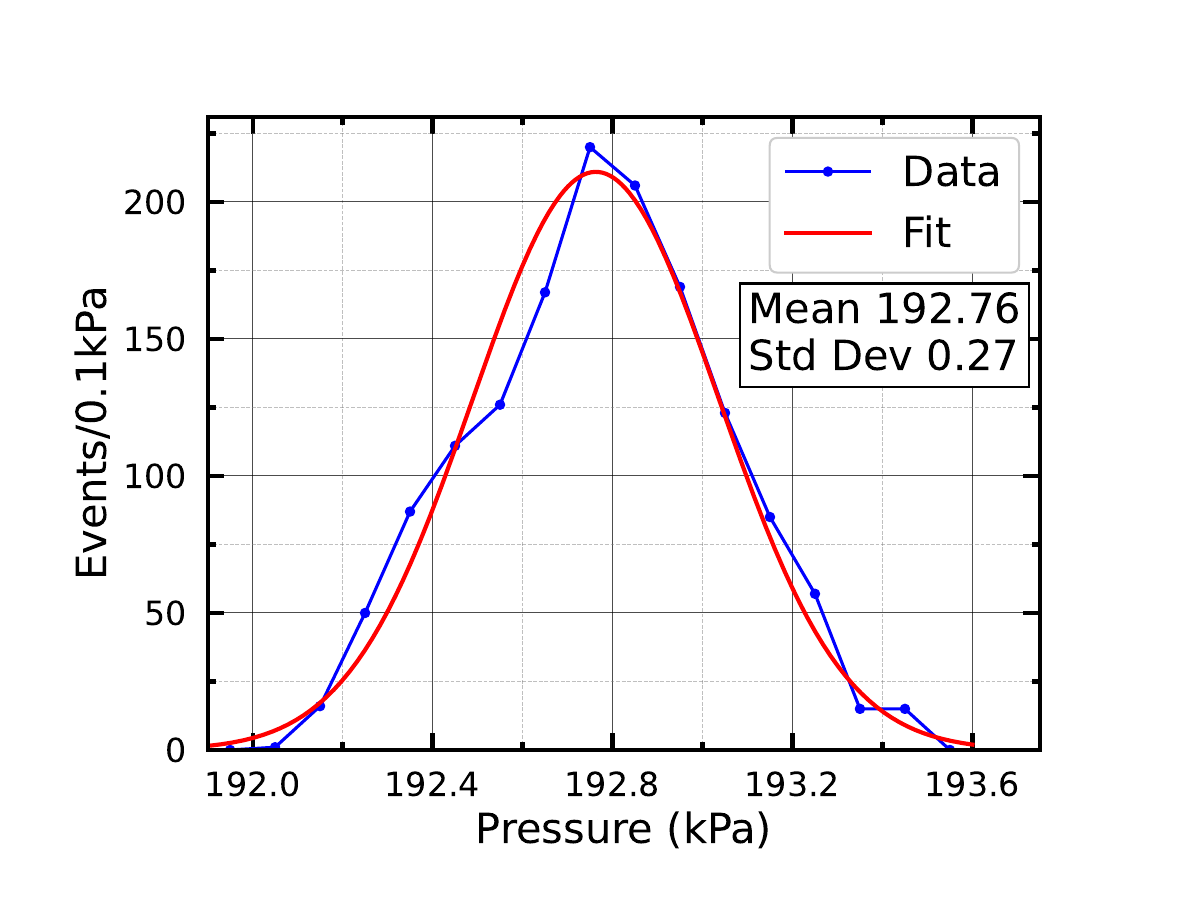}}
  \caption{Performance of the cooling towers during two cooling heads operate simultaneously. Shown are the fluctuation of cold finger temperature and inner pressure under two gas circulation and one all-liquid circulation.}
  \label{fig:p-2al600}
\end{figure}

In order to investigate the operational performance of the two cryocoolers running simultaneously, we powered on both units and set their cold head temperatures to 176~K. During this test, two gas circulation system were activated, each operating at a flow rate of approximately 80~slpm, along with a all-liquid circulation system running at about 780~kg/h (4.4~lpm). As shown in Figure~\ref{fig:p-2al600}, the combined heating power applied by the two heaters was approximately 1300~W, which stabilized the system pressure at $192.76\pm0.27$~kPa, indicating that the system still had 1300~W of available cooling capacity under these conditions. Furthermore, both cryocoolers successfully maintained the cold finger temperatures with fluctuations within $\pm$0.1~K, demonstrating excellent thermal stability and independent control.

Therefore, the new cryogenics system with two AL600 coldheads performs well during long-term operation and is capable of handling a large liquid xenon detector with a heat load exceeding 1500~W. 
At the same time, it maintains a pressure fluctuation of less than 1~kPa, satisfying the requirements of the PandaX-xT experiment.
This pressure fluctuation is larger than that of LZ (0.0558~kPa) and PandaX-4T (0.25~kPa) but smaller than that of XENONnT (2~kPa), assuming that LZ and XENONnT also used standard deviation as the metric.

\subsection{Performance of the emergency $LN_{2}$ cooler}

\begin{figure}
    \centering
    \includegraphics[width = 0.7\textwidth]{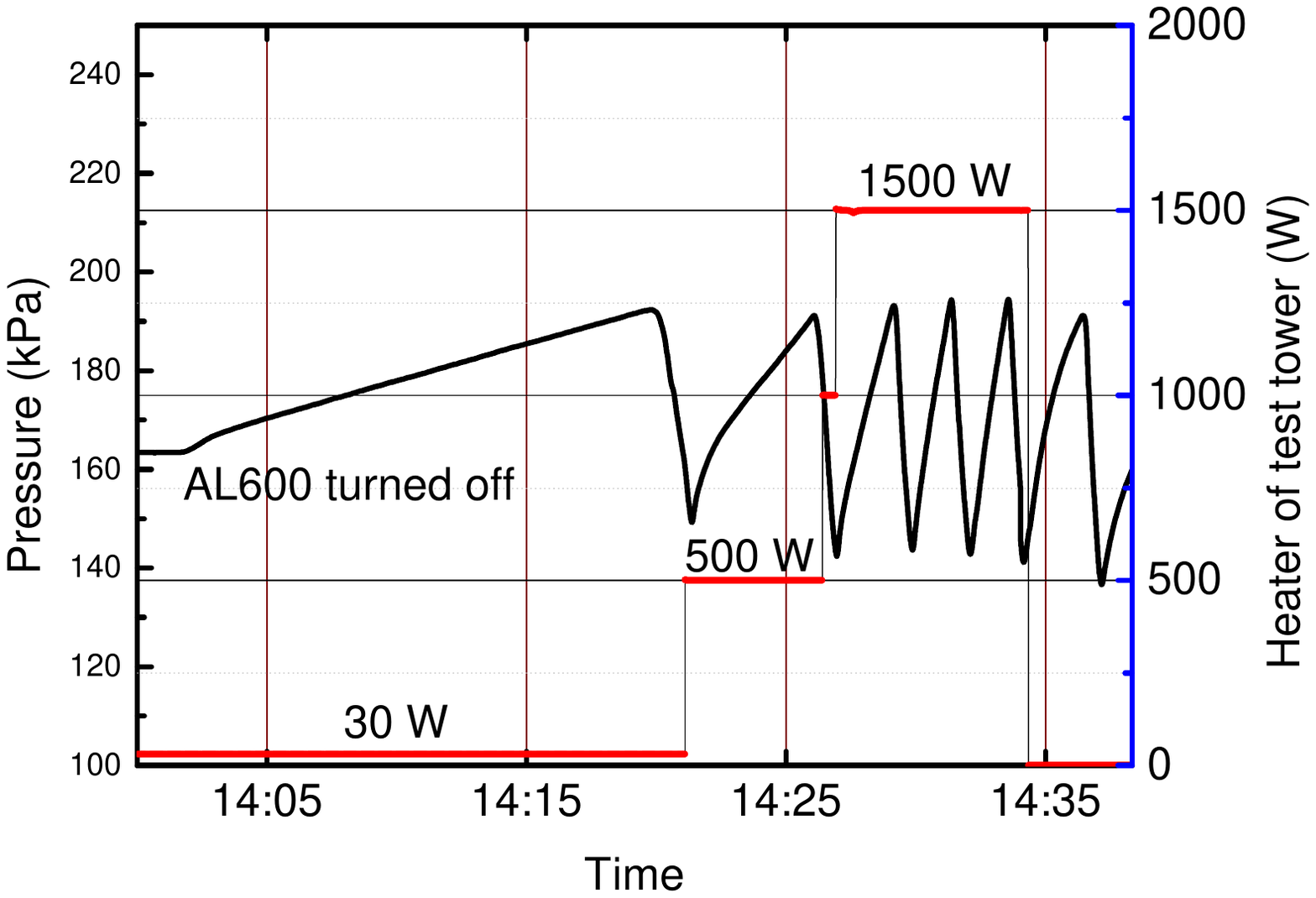}
    \caption{The test of $LN_{2}$ cooler using the test tower with about 15~kg liquid xenon. Red line indicates heater power, black line represents xenon pressure during the test. The pressure remains within the preset threshold under 1500~W heating power.}
    \label{fig:cooler-ln2}
\end{figure}

Before the 1-tonne liquid xenon detector was commissioned, a test was conducted using the test tower equipped with a heater to evaluate the emergency $LN_{2}$ cooler. Approximately 15~kg of xenon was first liquefied using the AL600 coldhead and then transferred into the inner liquid chamber of the test tower. The system was subsequently operated overnight to establish a stable thermal condition for the test. For safety during testing, the inner pressure was maintained at approximately 161~kPa using a 30 W heater. The upper and lower pressure thresholds for the $LN_{2}$ inlet valve were set to 190~kPa and 150~kPa, respectively. Subsequently, the AL600 cryocooler was turned off, and a series of tests were conducted at different heating powers (500~W and 1500~W) by heating the liquid xenon in the inner chamber of the test tower. The results are shown in Figure~\ref{fig:cooler-ln2}. As the heating power increased, the pressure rose more rapidly. At a heating power of 1500~W, the emergency 
$LN_{2}$ cooling system was still able to maintain stable operation, indicating a cooling capacity exceeding 1500~W at the liquid xenon temperature. 
This also confirms the validity of our theoretical design, as the measured cooling power of the liquid nitrogen coil exceeds 1500~W, which aligns well with the designed value of 1700~W.

Therefore, this new designed $LN_{2}$ cooler shows high cooling power, and it can handle emergency cases of unexpected power-off and malfunction of cryocoolers for future PandaX-xT experiment.

\section{Conclusion}
In this paper, the new cryogenics system prototype based on AL600 GM cryocoolers and $LN_{2}$ coil cooler has been constructed and experimentally investigated. The test results show that the prototype can provide a cooling capacity of approximately 1900~W at 178~K using two AL600 coldheads, along with an emergency $LN_{2}$ cooling capacity exceeding 1500~W. The coldheads operate stably at pressures of $212.95\pm0.31$~kPa and $210.61\pm0.35$~kPa, respectively, with each retaining an available cooling capacity greater than 650~W. Furthermore, the system exhibits high reliability and repeatability. In summary, it can be utilized for the next-generation large liquid xenon detector, PandaX-xT, in the future.

\acknowledgments
This project is supported in part by grants from National Key R\&D Program of China (Nos. 2023YFA1606200, 2023YFA1606201, 2023YFA1606202), National Science Foundation of China (Nos. 12090060, 12090061, 12090062, 12305121, U23B2070), Office of Science and Technology, Shanghai Municipal Government (Nos. 21TQ1400218, 22JC1410100, 23JC1410200, ZJ2023-ZD-003), and the Sichuan Science and Technology Program (Nos. 2024NSFSC1370). We thank for the support by the Fundamental Research Funds for the Central Universities. We also thank the sponsorship from the Chinese Academy of Sciences Center for Excellence in Particle Physics (CCEPP), Thomas and Linda Lau Family Foundation, New Cornerstone Science Foundation, Tencent Foundation in China, and Yangyang Development Fund. Finally, we thank the CJPL administration and the Yalong River Hydropower Development Company Ltd. for indispensable logistical support and other help. 

%\section*{References}
%\bibliographystyle{unsrt}
%\bibliography{mybibfile}

\end{document}